\documentclass[%
 reprint,
superscriptaddress,
 amsmath,amssymb,
 aps,
pra,
]{revtex4-2}
\usepackage{graphicx}

\usepackage{dcolumn}

\usepackage{hyperref}
\usepackage{xcolor}
\definecolor{cite}{rgb}{0.,0.,0.9}  
\hypersetup{colorlinks,linkcolor={cite},citecolor={cite},urlcolor={cite}}
\usepackage{gensymb}

\begin{document}


\title{Tailoring the Stability of a Two-Color, Two-Photon Rubidium Frequency Standard}
\author{Emily J. Ahern}
\affiliation{
 Institute for Photonics and Advanced Sensing (IPAS) and School of Physical Sciences, University of Adelaide, Adelaide SA 5005, Australia
}

\author{Sarah K. Scholten}
\email{sarah.scholten@adelaide.edu.au}
\affiliation{
 Institute for Photonics and Advanced Sensing (IPAS) and School of Physical Sciences, University of Adelaide, Adelaide SA 5005, Australia
}
\affiliation{
 ARC Centre of Excellence in Optical Microcombs for Breakthrough Science (COMBS)
}

\author{Clayton Locke}
\affiliation{
 Institute for Photonics and Advanced Sensing (IPAS) and School of Physical Sciences, University of Adelaide, Adelaide SA 5005, Australia
}
\affiliation{
QuantX Labs, Level 2, SpaceLab Building Frome Road, Lot Fourteen, Adelaide SA 5000
}

\author{Nicolas Bourbeau H\'{e}bert}
\affiliation{
 Institute for Photonics and Advanced Sensing (IPAS) and School of Physical Sciences, University of Adelaide, Adelaide SA 5005, Australia
}

\author{Benjamin White}
\affiliation{
 Institute for Photonics and Advanced Sensing (IPAS) and School of Physical Sciences, University of Adelaide, Adelaide SA 5005, Australia
}

\author{Andre N. Luiten}
\affiliation{
 Institute for Photonics and Advanced Sensing (IPAS) and School of Physical Sciences, University of Adelaide, Adelaide SA 5005, Australia
}
\affiliation{
 ARC Centre of Excellence in Optical Microcombs for Breakthrough Science (COMBS)
}
\affiliation{
QuantX Labs, Level 2, SpaceLab Building Frome Road, Lot Fourteen, Adelaide SA 5000
}

\author{Christopher Perrella}
\affiliation{
 Institute for Photonics and Advanced Sensing (IPAS) and School of Physical Sciences, University of Adelaide, Adelaide SA 5005, Australia
}
\affiliation{
 ARC Centre of Excellence in Optical Microcombs for Breakthrough Science (COMBS)
}
\affiliation{
 Centre of Light for Life and School of Biological Sciences, University of Adelaide, Adelaide, South Australia, 5005, Australia
}

\date{\today}

\begin{abstract}
Rubidium two-photon frequency standards are emerging as powerful contenders for compact, durable devices with exceptional stability.
The field has focused on single-color excitation to date.
Here we demonstrate the key advantages of a two-color excitation of a two-photon optical frequency standard based on the $5S_{1/2}\,{\rightarrow}\,5D_{5/2}$ transition of rubidium-87 utilising driving fields at 780\,nm and 776\,nm.
We show that utilising the $5P_{3/2}$ intermediate state to resonantly enhance the transition, we can for the first time attain frequency stabilities comparable to the rubidium single-color two-photon frequency standards, notably with approximately ten-fold less optical power and ten-fold lower rubidium vapor density.
Optimisation of the detuning from the $5P_{3/2}$ intermediate state, and optical powers of driving lasers, has a dramatic effect on the frequency stability, achieving the best short-term stability of any two-photon rubidium frequency standard to date, of $6\,{\times}\,10^{-14}$ at $\tau\,{=}\,1$\,s.
We demonstrate this level of performance is compatible with a compact geometry, by fully self-referencing the frequency standard using an integrated fiber frequency comb to simultaneously stabilize the 780\,nm laser’s detuning from the $5P_{3/2}$ intermediate state, and produce a frequency-stable microwave output.
A comprehensive noise characterization underpins our observations of this two-color frequency standard which explains the measured stability, showing this frequency standard is shot-noise limited initially before becoming limited by light shifts in the long-term.
This work represents a major advance towards a low size, weight, and power frequency standard based on this two-color excitation method.

\end{abstract}

\maketitle

\section{Introduction}

Atomic Frequency Standards have been in continuous development since the early 1950's~\cite{Ramsey_2005}, beginning with the development of a cesium-based atomic clock~\cite{essen1955atomic} and the subsequent redefinition of the second in 1967~\cite{lombardi2017historical}. 
Atomic clocks find applications in many areas from fundamental physics to everyday use in civilian position, navigation, and timing (PNT) services.
Atomic clocks are most commonly used for PNT services with microwave clocks being the traditional choice, with a research push towards optical clocks motivated by increased precision, and the potential for decreases in size, weight, and power consumption (SWaP).

Microwave atomic clocks have long been a popular choice as portable frequency standards since their first demonstration, and include the frequency standards used in PNT services~\cite{Mallette} such as global navigation satellite systems (GNSS)~\cite{Batori2020,Jaduszliwer2021,Schuldt2020,Hollberg2021}. 
Some archetypal examples of such portable frequency standards include the cesium frequency standard, the rubidium frequency standard~\cite{riley2019history}, and the passive hydrogen maser~\cite{gill2005optical}.
There has been significant effort focused on further improving the stability of portable microwave clocks. 
This includes a demonstration of a mercury trapped-ion atomic clock which operated continuously in low-Earth orbit for 2 months~\cite{Burt2021} which demonstrated a stability of approximately $2\,{\times}\,10^{-13}$ at an integration time $\tau\,{=}\,100$\,s, and $3\,{\times}\,10^{-15}$ at $\tau\,{=}\,23$\,days~\cite{Burt2021}. 
Another example is the microwave pulsed optical pumping (POP) rubidium clock that shows potential for use in GNSS~\cite{Jaduszliwer2021}, with a recent portable demonstration showing a stability of $2.6\,{\times}\,10^{-13}$ at $\tau\,{=}\,1$\,s and $2.3\,{\times}\,10^{-15}$ at $\tau\,{=}\,40,000$\,s (after drift removal)~\cite{Hao2024}.

The first atomic clocks based on optical transitions were demonstrated in 2001~\cite{ye2001,diddams2001} with development flourishing following the Nobel prize winning invention of the fully self-referenced optical frequency comb~\cite{Jones2000}.
This has enabled optical frequency standards to surpass microwave frequency standards in terms of their stability and systematic uncertainty. 
These optical frequency standards have achieved fractional frequency stabilities reaching into the $10^{-20}$ range over $\tau\,{=}\,10,000$\,s~\cite{Hutson2024}.
In addition to stability, optical atomic clocks have reached $10^{-19}$ systematic fractional frequency uncertainties~\cite{Brewer2019}.
Such optical frequency standards are confined to a laboratory setting, finding applications in measurement of the effects of general relativity and fundamental constants~\cite{Safronova2018}, and are integral for work towards the redefinition of the second~\cite{Katori2011,Bize2003,Reinhardt2007,Riehle2015,bregolin2017optical}. 

With recent advancements in decreasing the size and weight of lasers and other optical components, there has been a rapid acceleration in the development of compact and portable optical atomic clocks~\cite{Poli2014,Koller2017,Huang2020,Wang2020,Abbasov2020,Cao2022}.
A notable example is the iodine frequency standard that has been demonstrated aboard a sounding rocket with a flight time of 6\,minutes~\cite{Doringshoff2017,Doringshoff2019}.
In flight, this standard achieved a stability of $2\,{\times}\,10^{-13}$ at $\tau\,{=}\,100$\,s, while laboratory based iodine standards have shown a stability of $6\,{\times}\,10^{-15}$ at $\tau\,{=}\,1$\,s~\cite{Schuldt2017}.

Another promising option for a compact atomic frequency standard is the optical two-photon single-color rubidium frequency standard which has been demonstrated in recent years~\cite{Nez1993,Maurice2020,Newman2021,Lemke2022}.
This approach uses two-photon excitation at a wavelength of 778.1\,nm to excite the $5S_{1/2}\,{\rightarrow}\,5D_{5/2}$ transition in a thermal rubidium vapor.
This transition can be excited in a counter-propagating configuration, resulting in a narrow linewidth Doppler-free transition in a thermal vapor.
This has enabled a stability of $5\,{\times}\,10^{-15}$ (after drift removal) at $\tau\,{=}\,1$\,day to be reached~\cite{Lemke2022}. 

An alternative method of exciting this transition is a two-photon, two-color excitation scheme that uses one 780\,nm photon and one 776\,nm photon to drive the $5S_{1/2}\,{\rightarrow}\,5D_{5/2}$ transition~\cite{Perrella2019}.
This approach exploits the intermediate $5P_{3/2}$ state to resonantly enhance the transition rate~\cite{Bjorkholm1976}.
This allows for approximately ten-fold less optical power and ten-fold lower rubidium vapor density needed to achieve fractional frequency stability comparable to the single-color approach, achieving $1.5\,{\times}\,10^{-13}$ at $\tau\,{=}\,1$\,s~\cite{Perrella2019}.
A further key advantage lies in the flexibility to change the intermediate state detuning which has been explored to suppress light shifts~\cite{Gerginov2018, Hamilton2023}, as well as a theoretical study suggesting this flexibility can also be used to suppress temperature shifts~\cite{Nguyen2022}.
Therefore it would be instrumental to explore if this two-color excitation approach can exhibit equivalent stabilities to a direct two photon excitation and further, how we may simplify and enhance this approach to ultimately reach a goal of a compact instrument based on this premise.

In this article, we explore the benefits of the two-color excitation approach via two demonstrations. 
First, by exploiting intermediate state resonant enhancement, we demonstrate comparable frequency standard stabilities to previous single-color approaches.
Secondly, we optimise the detuning from the intermediate state together with the optical powers driving the transition to maximise the stability of the frequency standard.
These demonstrations are underpinned by a comprehensive analysis of noise sources.
We describe steps to simplify the apparatus with the overarching goal of a compact frequency standard.
To this end, we integrate an optical frequency comb into our system which provides down conversion from a stable optical frequency to microwave frequency output, and obviates the need for external frequency references making the system fully self-referenced. 
By implementing this self-referencing scheme we remove the complexity of an external source to frequency stabilize the 1560\,nm laser (such as by saturated absorption spectroscopy~\cite{Perrella2019}).
The presented frequency standard utilises predominantly fiber-coupled components in the telecommunications wavelength range allowing robust performance and reliable operation.
This represents a major step to a novel two-photon two-color rubidium frequency standard that holds promise for a compact portable device, with stable outputs in both the optical and microwave frequency domains.

\section{Two-Color Architecture}
For atomic clocks based on warm vapors such as the two-photon clock considered here, they are ideally limited by noise associated with the random arrival of photons at the detector, termed shot noise~\cite{Margolis2010}.
The shot noise sets the signal-to-noise ratio (SNR) of the measured atomic transition with linewidth ($\Gamma_e$) and transition frequency ($\omega$).
This results in a frequency standard stability, $\sigma(\tau)$, at an integration time, $\tau$, of~\cite{Margolis2010}:
\begin{equation}
    \sigma(\tau) \propto \frac{\Gamma_e}{\omega} \frac{1}{\text{SNR}} \frac{1}{\sqrt{\tau}},
\end{equation}
from which a shot-noise limited frequency standard will improve its stability as $1/\sqrt{\tau}$. 

Shot-noise and thus SNR are related to the excitation rate ($W$) of the observed atomic transition, and the accompanying number of photons detected.
In a two-photon excitation scheme, the on-resonance transition rate depends on laser detuning from an intermediate state ($\Delta_i$) and intensity of the two driving lasers ($I_1$ and $I_2$), which can be expressed as~\cite{Bjorkholm1976, Hamilton2023}: 
\begin{equation}\label{eqn:TransRate}
    W \propto \frac{N}{\Gamma_e}\frac{I_1 I_2}{4 \Delta_i^2 + \Gamma_i^2}
\end{equation}
where $N$ is the atomic number density and $\Gamma_i$ and $\Gamma_e$ are the linewidth of the ground to intermediate state transition, and intermediate to excited state transition respectively.
In the case of the $5S_{1/2}\,{\rightarrow}\,5D_{5/2}$ two-photon transition, the intermediate state is the $5P_{3/2}$ state and the excited state is the $5D_{5/2}$ state.
Excitation of the two-photon transition is measured by detection of fluorescence produced at 420\,nm via the $6P_{3/2}$ decay path (Fig.~\ref{fig:ErrorSignal}(a)) in order to measure the strength of the transition~\cite{Perrella2019}.

In the single-color excitation scheme, two 778.1\,nm photons from a single laser drive the transition such that $I_1\,{=}\,I_2$.
The counter-propagating excitation produces a Doppler-free linewidth of $\Gamma_e\,{=}\,667$\,kHz limited by the lifetime of the excited state~\cite{Sheng2008}.
The large detuning from the intermediate state, $\Delta_i\,{=}\, 1.1$\,THz, results in a weak transition, with laser powers of approximately 30\,mW required to drive the transition with 1/$e^2$ intensity beam radius of 0.66\,mm giving $I_1\,{=}\,I_2\,{\approx}\,22$\,mW/mm$^2$~\cite{Martin2018}.

In the two-color scheme, the intermediate state detuning is decreased to $\Delta_i\,{\approx}\,1.5$\,GHz, increasing the two-photon transition rate. 
However, this change also introduces residual Doppler broadening of approximately $\Gamma_e = 4$\,MHz~\cite{Bjorkholm1976, Perrella2019}.
As a result of decreasing the intermediate state detuning, similar driving optical powers leads to a factor of $10^5$ increase in the excitation rate of the two-color configuration when compared to the single-color scheme.
Alternatively, the optical powers can both be reduced by a factor of approximately $300$ times for the two-color configuration to achieve a similar excitation rate to the single-color scheme.
The two-color excitation scheme provides the flexibility to tune the intermediate state detuning, as well as the power of each laser independently.
This allows tailoring of the frequency standard's stability, trading off intermediate state detuning against optical power of the lasers, for example if available optical power is limited, small intermediate state detuning can achieve similar results to high-power far detuned situations.

\section{Experiment}
\label{sec:experiment}
\begin{figure}[t]
    \includegraphics[width=9cm]{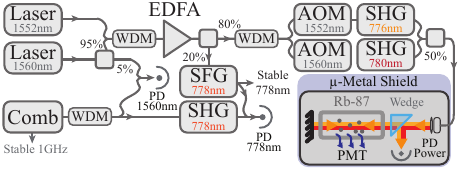}
    \caption{The experimental setup. Optical components and fibers are shown in grey with solid lines, free-space laser beams are shown in orange and red solid lines. WDM: Wavelength Division Multiplexer; EDFA: Erbium Doped Fiber Amplifier; SFG: Sum-Frequency Generator; SHG: Second Harmonic Generator; AOM: Acousto-Optic Modulator; PMT: Photomultiplier Tube; PD: Photodetector.}
    \label{fig:ExperimentalDiagram}
\end{figure}

\begin{figure}[t]
    \includegraphics[width=\columnwidth]{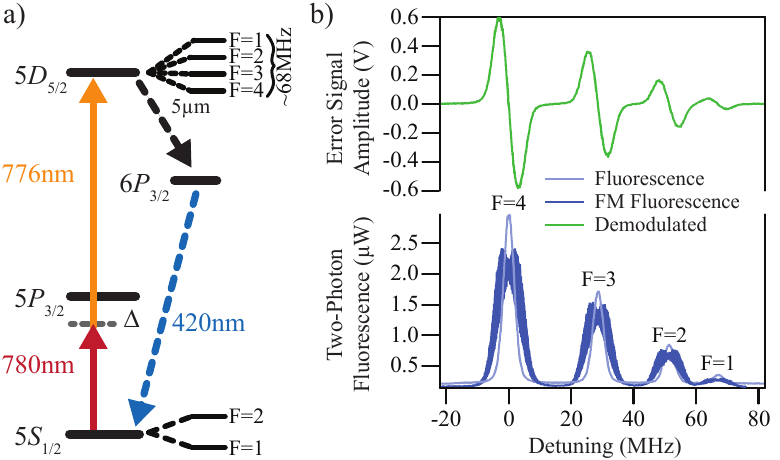}
    \caption{a) Two-color excitation scheme of the $5S_{1/2}\rightarrow5D_{5/2}$ transition in the two-photon frequency standard. One 780\,nm and one 776\,nm photon excite the transition via an intermediate state $5P_{3/2}$, with successful excitation evidenced by the observation of 420\,nm fluorescence via the $6P_{3/2}$ state decay pathway. Detuning from the intermediate state ($\Delta_i$) is set to 1.5\,GHz. b) Bottom: PMT-detected 420\,nm fluorescence (light blue) from atomic decay through the $6P_{3/2}\,{\rightarrow}\,5S_{1/2}$ transition. Four peaks arise from the hyperfine splitting of the $5D_{5/2}$ energy level. Frequency modulation of $f_{mod}$ is applied to the detected fluorescence (dark blue). Top: Frequency discriminator/error signal (green) used to lock the 1552\,nm laser frequency to the two-photon transition, derived from the frequency-modulated (FM) fluorescence.}
    \label{fig:ErrorSignal}
\end{figure}

The two-color two-photon rubidium frequency standard presented here can be broadly split into two sections: the laser preparation stage, and the physics package which holds the rubidium cell in which the two-photon transition is excited (Fig.\,\ref{fig:ExperimentalDiagram}).
Two fiber lasers at 1552\,nm and 1560\,nm are frequency-doubled and used to drive the two-color transition.
The 1552\,nm and 1560\,nm lasers are combined in a wavelength division multiplexer (WDM) and amplified by an erbium-doped fiber amplifier (EDFA). 
The light is passed through a 80:20 splitter, with 80\% of the light directed into a WDM that separates the two laser wavelengths, with each wavelength directed through its own acousto-optic modulator (AOM) and a second harmonic generation crystal (SHG). 
These SHGs convert the 1552\,nm and 1560\,nm light into the 776\,nm and 780\,nm light respectively, which are required to excite $5S_{1/2}\,{\rightarrow}\,5D_{5/2}$ transition (Fig.\,\ref{fig:ErrorSignal}a). 
Both colors are combined via a 50:50 fiber combiner and are co-linearly launched into the physics package, within which the light passes through a polarizer and optical wedge before entering the rubidium vapor cell. 
Using a mirror at the rear of the cell, the light is retro-reflected, allowing us to perform counter-propagating excitation, resulting in a decrease of the effect of Doppler broadening.
The resulting two-photon lineshape is a convolution of the natural line shape (Lorentzian) and Doppler broadening (Gaussian), forming a Voigt spectral profile with a line width of approximately $\Gamma_e\,{=}\,3.47\,{\pm}\,0.02$\,MHz~\cite{Perrella2019, Bjorkholm1976}. 
Excitation of the two-photon transition is measured via 420\,nm fluorescence, produced via decay from $5D_{5/2}\,{\rightarrow}\,5S_{1/2}$ through the $6P_{3/2}$ state. 
Approximately 8\% of the decay from the excited state will follow this path~\cite{Safronova2004, Hamilton2023, Hassanin2023}. 
The 420\,nm fluorescence is detected by two photomultiplier tubes (PMTs) mounted next to the cell within the physics package. 
Multiple PMTs are used to capture as much fluorescence as possible from the rubidium cell.

We heat the rubidium cell to an operating temperature of between 60\,$\degree$C and 70\,$\degree$C, increasing the density of rubidium atoms and thus, the amount of fluorescence produced. 
This is achieved by embedding heater cartridges within a large thermal heat capacity stainless steel oven encapsulating the rubidium cell. 
The temperature of the rubidium cell is stabilized using a combination of passive isolation and active control. 
Passive isolation is achieved via the stainless steel oven, a layer of 3D printed Ultem (polyetherimide) insulation, and a surrounding $\mu$-metal shield. 
This isolates the vapor cell from fluctuations in environmental temperature and slows the loss of heat, reducing the power requirement of heating the vapor cell.
The passive thermal isolation provides a 500-fold reduction to ambient temperature fluctuations with a time-constant of 12\,ks. 
The temperature of the stainless-steel housing is sensed via a 4-wire measurement of a 10\,k$\Omega$ Negative Temperature Coefficient (NTC) thermistor which is used to actively control the temperature of the stainless-steel housing via the four heater cartridges in a feedback loop. 
An out-of-loop 4-wire 10\,k$\Omega$ NTC thermistor is used to independently monitor the temperature of the housing which shows 2\,mK fluctuations over 1200\,s.

The 1552\,nm laser is frequency stabilized by using frequency modulation spectroscopy. 
We apply a frequency modulation, $f_{mod}$, to the 1552\,nm laser with $f_{mod}$ and depth similar to that demonstrated in Ref.~\cite{Perrella2019}.     
This is detected as an amplitude modulation on the 420\,nm fluorescence and demodulated to produce an error signal with the zero crossing of the $F\,{=}\,4$ hyperfine feature (Fig.~\ref{fig:ErrorSignal}a) used as the lock point to stabilize the 1552\,nm laser frequency. 

The nominal optical powers of the 780\,nm and 776\,nm lasers at the rubidium cell are on the order of 1\,mW each.
The power of each laser is stabilized by detecting a fraction of the lasers' power as close as possible to the rubidium cell.
Two separate photo-diodes (PDs) are used for power control loops and diagnostic out-of-loop measurements.
Measuring immediately before the vapor cell means that both colors are spatially overlapped, and the power of each laser must be separated in either the  spatial or temporal domain.
Spatial separation would require free-space propagation over long distances~\cite{Perrella2019}, which would be impractical for this purpose.
Thus, temporal separation is utilised which is implemented by square-wave amplitude modulation of the 780\,nm light at a frequency of $f_{AC}\,{\ll}\,f_{mod}$, applied via the 1560\,nm AOM.
The square-wave signal is detected on the PD (PD Power in Fig.\,\ref{fig:ExperimentalDiagram}) and used in a proportional integral (PI) filter to feedback to the AOM for power control, thus actively controlling the power of the 780\,nm beam.
The 776\,nm power control, in contrast, is a DC stabilization loop that drives the 1552\,nm AOM to maintain the DC level of the same PD at a given set point.

The use of an AOM to apply frequency modulation at $f_{mod}$ to the 1552\,nm laser for frequency stabilization to the two-photon transition creates an unwanted laser power modulation at the same frequency, referred to as residual amplitude modulation (RAM). 
This leads to an unwanted amplitude modulation of the 420\,nm fluorescence at $f_{mod}$ which feeds into the frequency control loop for the 1552\,nm laser and thus the frequency standard’s output frequency.
To suppress the RAM, the in-phase and quadrature components of the RAM on the 776\,nm laser power are detected on the power control PD. 
A tone at $f_{mod}$ is applied to the AOM RF drive amplitude with the same amplitude but opposite phase as the detected RAM on the power control PD (PD Power in Fig.\,\ref{fig:ExperimentalDiagram}), thus cancelling the RAM on the 776\,nm laser power. 

When locked, the 1552\,nm and 1560\,nm laser frequencies are referenced to the rubidium two-photon transition making the sum of the two laser frequencies stable.
To derive a frequency stable optical output from the frequency standard, 20\% of the light is separated after the first WDM and is directed through a sum frequency generator (SFG). 
This generates a frequency stable 778\,nm optical output by summing the frequencies of the 1552\,nm and 1560\,nm lasers.
To down-convert this stability to the microwave domain, we utilise a home-built fiber frequency comb similar to the design developed at NIST~\cite{Sinclair2015InvitedAA}. 
The repetition rate, $f_{\text{rep}}$, of the comb (200\,MHz) is stabilized to the frequency stable optical output of the frequency standard at 778\,nm via a phase-locked loop, measured via PD 778\,nm in Fig.\,\ref{fig:ExperimentalDiagram}. 
The frequency comb's Carrier-Envelope Offset (CEO) is stabilized via the traditional $f$-$2f$ self referencing scheme.

The intensity and detuning of the two lasers from the intermediate state couple to the atomic energy levels creating a shift in their energy, an effect known as light shifts (discussed in further detail in Sec.~\ref{sec:LightShift}). 
If either the intensity or detuning of the lasers fluctuate, it will generate a fluctuation in light shifts and instability in the frequency standard’s output frequency.
The power stabilization of both lasers reduces the creation of fluctuating light shifts via this mechanism. 
We stabilize the intermediate state detuning ($\Delta_i$) in a compact manner by utilising the stabilized frequency comb by frequency stabilizing the 1560nm laser to a mode of the comb via a phase-locked loop with the optical beat note between the laser and comb detected on PD 1560\,nm in Fig.\,\ref{fig:ExperimentalDiagram}.
This locking scheme introduces an interaction between locking loops controlling the frequency of both fiber lasers and the repetition rate of the frequency comb. 
This interaction is generated because frequency instability in the 1560\,nm laser frequency lock is imparted onto the atomic transition via light shifts. 
In turn, this introduces an instability in the frequency lock of the 1552\,nm laser, which is locked to the two-photon transition.
The instability is then transferred to the repetition rate of the frequency comb and back to the 1560\,nm laser frequency.
Therefore, excessive gain in the 1560\,nm frequency lock will lead to oscillations of the 1552\,nm lock and repetition rate frequency lock. 
Oscillations are avoided by ensuring that the coupling between locking loops has a gain less than 1, thus any oscillations are attenuated rather than amplified.
In section~\ref{sec:LightShift}, we characterize the frequency standard’s sensitivity to the intermediate state detuning (by varying the 1560\,nm frequency) to be 40\,Hz/MHz.
This coupled with a low gain 1560nm frequency lock to the comb ensures stable operation of the frequency standard.

The stability of the frequency standard is measured in the microwave domain by comparison to a lab-based cryogenic sapphire oscillator (CSO)~\cite{Tobar2006}.
This is achieved by comparing the optical frequency comb's fifth harmonic of the repetition rate, $5{\times}f_\text{rep}\,{=}\,1$\,GHz, with a 1\,GHz output from the CSO, resulting in a beat note in the ten's of kilohertz which is measured on a commercial frequency counter.
In addition to this, we simultaneously measure the out-of-loop optical powers (780\,nm AC, 780\,nm and 776\,nm DC powers, 776\,nm RAM), along with the cell temperature and average 420\,nm fluorescence power ($P_{420}$) to enable calculation of pressure shifts and shot noise for a thorough noise source analysis.

\section{Results and Noise Analysis} 

\begin{figure}[t]
    \includegraphics[width=\columnwidth]{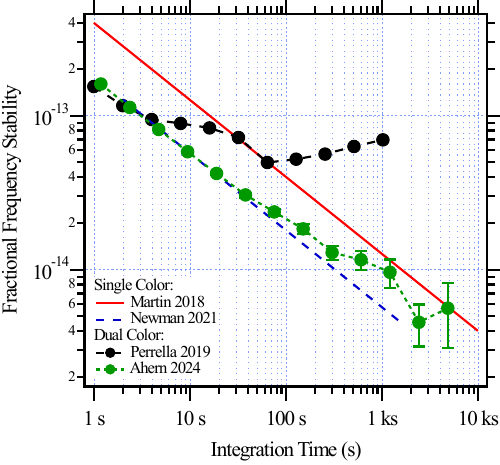}
    \caption{Comparison of two-photon frequency standard stabilities.
    Two single-color frequency standard (drift removed) fractional frequency stabilities are presented~\cite{Martin2018,Newman2021}, along with the previously presented stability of this two-color approach~\cite{Perrella2019}. 
    We see that the two-color approach presented here yields a fractional frequency stability (calculated by subtracting residual 780\,nm light shifts) is on par with the single color design, and a significant improvement on previous results.}
    \label{fig:FSCompare}
\end{figure}

\label{sec:Results}

\begin{figure}[t]
    \includegraphics[width=\columnwidth]{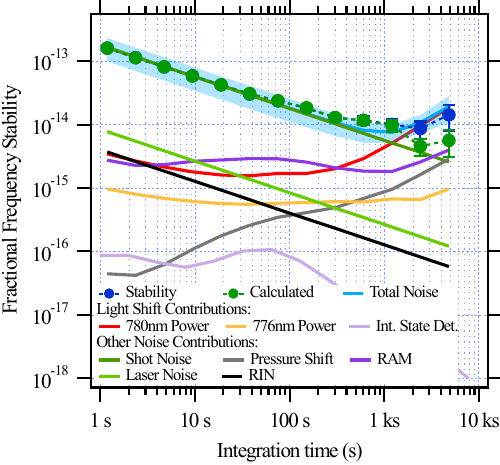}
    \caption{Measured fractional frequency stability of the two-color frequency standard (closed blue circles, dashed line) along with calculated stability by subtracting residual 780\,nm light shifts (closed green circles, dashed line).
    Light shift contributions to the frequency stability are also shown: 780\,nm light shift (red), 776\,nm light shift (orange), intermediate state detuning (lavender).
    Additional noise contributions are also shown: shot noise (dark green), pressure shift (grey), RAM (purple), laser frequency noise (light green), RIN (black).
    The summation (Total Noise) of all noise sources is shown to agree with measurements (light blue). 
    Error on the total noise line is given by the quadrature sum of the error on all contributing noise traces. 
    Note that the Zeeman noise contribution is below the range of this graph with a contribution of $3.2\,{\times}\,10^{-20}$ at $\tau\,{=}\,1200$\,s.} 
    \label{fig:totalSRAV}
\end{figure}

In this section we present the stability results of the two-color two-photon frequency standard and associated noise analysis.
The frequency stability of the frequency standard is presented in Figs.~\ref{fig:FSCompare} and~\ref{fig:totalSRAV} which demonstrates a fractional frequency stability of $1.6\,{\times}\,10^{-13}$ at $\tau\,{=}\,1$\,s and averages down to $8.6\,{\times}\,10^{-15}$ at $\tau\,{=}\,2400$\,s.
By subtracting residual 780\,nm optical power fluctuations from the stable frequency standard output, we achieved a calculated frequency standard stability of $4.5\,{\times}\,10^{-15}$ at $\tau\,{=}\,2400$\,s.
This is a significant improvement to the previously presented two-color two-photon frequency standard stability~\cite{Perrella2019}, as shown in Fig.~\ref{fig:FSCompare}.
Furthermore, the results are comparable to single-color two-photon frequency standard configurations~\cite{Martin2018, Newman2021}, also presented in Fig.~\ref{fig:FSCompare}. 
Compared to the single-color approach, we achieve these results with between two and ten times less optical power interacting with the rubidium vapor, and an approximately $30-40\,\degree$C cooler vapor cell, corresponding to an order of magnitude lower rubidium vapor density~\cite{Steck2023}.
This highlights the benefits of the two-color approach in strongly driving the two-photon transition.

We now discuss the noise sources affecting the stability of the two-color frequency standard with the following subsections detailing the individual noise contributions. 
We separate the noise contributions into two categories: those that cannot be measured whilst the frequency standard is operational, and those that can. 
We first discuss the noise sources that cannot be measured whilst the frequency standard is in operation, which are: shot noise, relative intensity noise (RIN), laser frequency noise at the modulation frequency, and Zeeman shifts of the two-photon transition. These are unable to be measured during operation because the measurement of each in some way perturbs the operation of the frequency standard.
Whilst the frequency standard is in operation we measure noise contributions from light shifts, pressure shifts, and residual amplitude modulation (RAM). 
The frequency standard's sensitivity to these noise sources, along with their stability contributions, are summarised in Table~\ref{noisesources}.
We combine the noise sources to predict the frequency standard's stability at varying integration times which is presented in Fig.~\ref{fig:totalSRAV} along with the measured/calculated stability contribution of each noise source.

\begin{center}
    \begin{table}[b]
    \begin{tabular}{l c c c}
        \hline \hline
        Source                      & Sensitivity               & Stability                             & Stability                     \\
                                    &                           & at $\tau\,{=}\,1200$\,s               & Contribution                  \\
                                    &                           &                                       & at $\tau\,{=}\,1200$\,s       \\ 
        \hline 
        Shot Noise                  & 20\,Hz/$\mu$V             & 99\,nV                                & $5.2\,{\times}\,10^{-15}$     \\
        Light Shifts:               &                           &                                       &                               \\
        \quad780\,nm Power          & -32.7\,Hz/$\mu$W          & 60\,nW                                & $5.1\,{\times}\,10^{-15}$     \\
        \quad776\,nm Power          & 2.3\,Hz/$\mu$W            & 110\,nW                               & $6.7\,{\times}\,10^{-16}$     \\
        \quad Int. State Det.       & 190\,Hz/MHz               & 20\,Hz                                & $8.5\,{\times}\,10^{-18}$     \\
        RAM                         & 4.3\,Hz/$\mu$W            & 0.16\,$\mu$W                          & $1.8\,{\times}\,10^{-15}$     \\
        Pressure Shift              & -460\,Hz/K                & 0.8\,mK                               & $9.5\,{\times}\,10^{-16}$     \\
        Freq. Noise$^\text{a}$      & -                         & 0.094\,Hz                             & $2.4\,{\times}\,10^{-16}$     \\
        RIN$^\text{a}$              & 1.3\,MHz                  & $3.6\,{\times}\,10^{-8}$              & $1.2\,{\times}\,10^{-16}$     \\
        Zeeman Shift                & 114\,Hz/${\text{G$^2$}}$* & 3\,$\mu$G                             & $3.2\,{\times}\,10^{-20}$     \\
        \hline \hline
    \end{tabular}
    \caption{A summary of noise sources and their contribution to the frequency stability of the frequency standard at $\tau\,{=}\,1200$\,s. 
    *Value taken from Ref.~\cite{Martin2018}. 
    $^\text{a}$Quadrature sum of contributions from both lasers.}
    \label{noisesources}
    \end{table}
\end{center}

\subsection{Shot Noise}
\label{sec:ShotNoise}
Shot noise originates from the discrete and random arrival of 420\,nm fluorescence photons at the photo-multiplier tubes (PMTs). 
This directly contributes to the signal-to-noise ratio (SNR) of the two-photon fluorescence measurement to which the frequency standard is stabilized, thus directly affecting the frequency standard's stability. 
The resulting photo-current shot noise on the photo-cathode is given by a power spectral density  
of the form
\begin{equation}    \label{eqn:shotnoise}
    S(f)_I = \sqrt{2 q I}
\end{equation}
where $q$ is the elementary charge, and $I$ is the photo-cathode current.
 
Taking into account the gain factors of the PMTs and analog electronics, $G$, we calculate the shot noise level at the input to the Field Programmable Gate Array (FPGA) that stabilizes the 776\,nm laser to the two-photon transition.
Using the frequency discriminator, $\delta_\textrm{error}$, shown in Fig.\,\ref{fig:ErrorSignal}, the shot noise power spectral density in frequency units, calculate via $S(f)_{\textrm{Hz}}\,{=}\,S(f)_I G/\delta_\textrm{error}$, 
was converted into an Allan deviation, $\sigma(\tau)$, with~\cite{Rutman1991, Turner2002}:
\begin{equation}
    \label{eqn:PSDintegral}
    \sigma(\tau)^2=2  \int_{0}^{\infty} \,S(f)_{\textrm{Hz}}^2 \frac{\sin(\pi f \tau)^4}{(\pi f \tau)^2}\,\text{d}f
\end{equation}
where $f$ is the Fourier frequency of the noise spectrum and $\tau$ is the integration time of the stability measurement. 
We find a contribution of $1.8\,{\times}\,10^{-13}/\sqrt{\tau}$ for shot noise, matching our measured frequency standard stability over $1\,{<}\,{\tau}\,{<}\,500$\,s, as seen in Fig.~\ref{fig:totalSRAV}. 
This demonstrates that over these time scales our frequency standard is shot noise limited.

\begin{figure}[t]
    \includegraphics[width=\columnwidth]{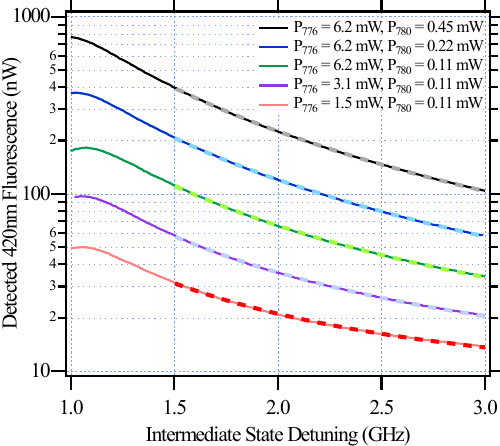}
    \caption{As detuning from the intermediate state increases the strength of the transition decreases. 
    Here this is demonstrated via measurement of the 420\,nm fluorescence for different optical powers of the 780\,nm and 776\,nm lasers driving the transition.
    Dashed lines are fits to Eqn.~\ref{eqn:TransRate} illustrating agreement with theoretical expectations.}
    \label{fig:Power_Detuning}
\end{figure}

The two-color approach allows us to tailor the two-photon transition rate, $W$, via laser powers and intermediate state detuning, as described by Eqn.~\ref{eqn:TransRate}.
To verify this relationship, we measure the 420\,nm fluorescence with different power combinations over a range of intermediate state detunings as shown in Fig.~\ref{fig:Power_Detuning}.
As expected, from Eqn.~\ref{eqn:TransRate}, a decrease in detected fluorescence is observed with increased detuning from the intermediate state, with fits to Eqn.~\ref{eqn:TransRate} showing good agreement with the data between 1.5\,GHz and 3\,GHz giving an average full-width half maximum (FWHM) of $\Gamma_i\,{=}\,0.34\,{\pm}\,0.11$\,GHz, a result of Doppler broadening of the intermediate state.
As the intermediate state detuning is reduced, the fluorescence starts to turn over and decrease at approximately 1\,GHz detuning.
This is due to increased absorption of the 780\,nm beam from the single-photon $5S_{1/2}\,{\rightarrow}\,5P_{3/2}$ transition, attenuating the 780\,nm laser power as it passes through the rubidium cell, thus decreasing the two-photon transition rate and fluorescence produced in the region observed by the PMTs.
We also observe that the PMT fluorescence for any given detuning is proportional to the product of the driving laser powers ($P_{776}\,{\times}\,P_{780}$) as expected from Eqn.~\ref{eqn:TransRate}. 
This is seen in Fig.~\ref{fig:Power_Detuning} by consecutive traces from top to bottom being approximately a factor of two different in their detected fluorescence which corresponds to the decrease in the product of the driving laser powers.

\begin{figure}[t]
    \includegraphics[width=\columnwidth]{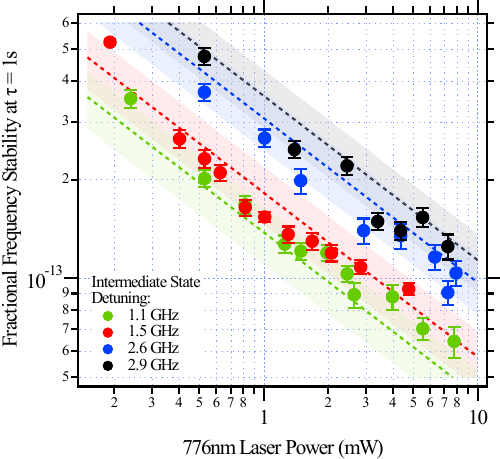}
    \caption{Fractional frequency stability of the frequency standard at $\tau\,{=}\,1$\,s at various 776\,nm powers and intermediate state detunings. The change in fractional frequency stability is a result of changing 420\,nm fluorescence production and changing shot noise contributions. 
    Dashed lines are predicted stability's based on our shot noise measurement and known behaviour of the two-photon transition rate, Eqns.~\ref{eqn:TransRate} and \ref{eqn:shotnoise}. 
    Shaded regions around the dashed lines indicate uncertainty in the shot noise measurement.}
    \label{fig:SRAV776Power}
\end{figure}

As seen by Fig.~\ref{fig:Power_Detuning}, the two-color approach allows tailoring of the two-photon transition rate, $W$, as described by Eqn.~\ref{eqn:TransRate}.
As the photo-current of the PMTs is proportional to $W$, we can tune the shot noise contribution to the frequency stability of the standard, $S(f)_I$, as described by Eqn.~\ref{eqn:shotnoise}, by utilising individual laser powers and detuning from the intermediate state.
We have demonstrated that the frequency standard is shot noise limited between $1\,{<}\,{\tau}\,{<}\,500$\,s, as seen in Fig.~\ref{fig:totalSRAV}, thus tailoring $W$ gives us the ability to tune the stability of the frequency standard within this integration time window.
Figure~\ref{fig:SRAV776Power} illustrates this effect, in which the fractional frequency stability at $\tau\,{=}\,1$\,s is presented for various intermediate state detunings and 776\,nm optical powers.
As the 776\,nm power is increased, we see a corresponding improvement in the fractional frequency stability at $\tau\,{=}\,1$\,s due to the increased 420\,nm fluorescence detected.
Similarly, as the intermediate state detuning is decreased, we observe a improvement in stability due to an increase in 420\,nm fluorescence, as seen in Fig.~\ref{fig:SRAV776Power}.
Dashed lines on Fig.~\ref{fig:SRAV776Power} are the predicted stabilities from our measured shot noise level, scaled by the known behaviour of the two-photon transition rate and shot noise as a function of detuning and power, described by Eqns.~\ref{eqn:TransRate} and \ref{eqn:shotnoise}.
Shaded regions around the dashed lines indicate uncertainty in the shot noise measurement.
By reducing the intermediate state detuning to 1.1\,GHz and tuning the optical power of both the 780\,nm and 776\,nm respectively, we maximise the two-photon transition rate and achieve a fractional frequency stability of $6.4\,{\times}\,10^{-14}$ at $\tau\,{=}\,1$\,s.

\begin{figure}[t]
    \includegraphics[width=\columnwidth]{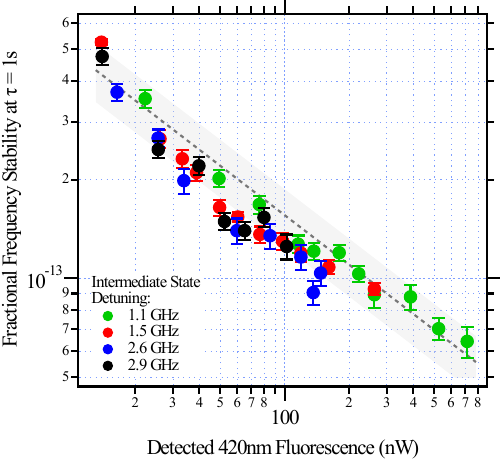}
    \caption{Varying the intermediate state detuning changes the amount of fluorescence produced from the two-photon transition. A result of this is an ability to decrease our shot noise, thereby increasing the 1\,s stability, by moving closer to the intermediate state. 
    Dashed line is the predicted stability based on shot noise measurements and scaled with Eqn.~\ref{eqn:shotnoise}. 
    Shaded region around the dashed line indicates uncertianty in the shot noise measurement.
    }
    \label{fig:SRAVPMTDC}
\end{figure}

Plotting each of the stabilities presented in Fig.~\ref{fig:SRAV776Power}, measured with different intermediate state detuning and 776\,nm optical power, against the 420\,nm fluorescence detected by the PMTs shows that all the data collapses onto a single line as shown in Fig.~\ref{fig:SRAVPMTDC}.
This demonstrates the shot noise limited nature of the frequency standard over this entire parameter space.
Again, the dashed line and shaded area are the predicted stabilities based on our shot noise measurement its uncertainty respectively.

It can be seen from Figs.~\ref{fig:SRAV776Power} and \ref{fig:SRAVPMTDC} that a given frequency standard stability can be attained with different combinations of intermediate state detunings and 776\,nm optical powers.
This demonstrates the flexibility in operational conditions for the two-color excitation scheme, where we can utilise a small intermediate state detuning and low 780\,nm and 776\,nm optical powers to achieve a comparable stability to a far-detuned, high power configuration.

\subsection{Relative Intensity Noise}
Relative intensity noise (RIN) of the 1552\,nm and 1560\,nm lasers is the fractional noise in the power of each laser. 
Laser power fluctuations are suppressed within the bandwidth of the power control loops (1\,kHz) however, higher frequency power fluctuations remain uncontrolled. 
RIN of either laser at the $f_{mod}$ for the two-photon frequency lock leads to noise in the zero crossing (lock point) of the frequency discriminator, see Fig.~\ref{fig:ErrorSignal}, used to stabilize the 1552\,nm laser frequency.

For both the 1552\,nm and 1560\,nm lasers we measured the power spectral density (PSD) of the optical power, $P(f)$, and average optical power, $\overline{P}$, at the relevant operating conditions. 
The RIN of each laser was then calculated using $\text{RIN}(f)\,{=}\,20 \log_{10} \left( P(f)/\overline{P}\right)$.
The RIN PSD of each laser was measured to be $-112\,\text{dBc}/\sqrt{\text{Hz}}$ or in fractional units $2.5\,{\times}\,10^{-6}/\sqrt{\text{Hz}}$ at our modulation frequency of $f_{mod}$.
To calculate the noise contribution of the RIN to the frequency standard's fractional frequency stability, we calculate its contribution to the two-photon frequency lock.  
The frequency noise contribution of the RIN is calculated via $S(f)_{\textrm{Hz}}\,{=}\,\text{RIN}(f)\times\textrm{D}/\delta_\textrm{error}$,
where the second term applies the RIN to the amplitude of the 420\,nm fluorescence signal, D, and converts to frequency noise by dividing by the frequency discriminator $\delta_\textrm{error}$.
This gives a value of $S(f_{mod})_{\textrm{Hz}}\,{=}\,3.1\,\text{Hz}/\sqrt{\text{Hz}}$ per laser. 
Using Eqn.~\ref{eqn:PSDintegral} we convert this to fractional frequency units giving a noise contribution of $4.0\,{\times}\,10^{-15}/\sqrt{\tau}$.

\subsection{Laser Frequency Noise}
Similar to RIN described above, the 1552\,nm and 1560\,nm lasers possess frequency noise, some of which is controlled by their respective frequency control loops, whilst frequency noise outside the bandwidth of these control loops is uncontrolled.
Similar to RIN, laser frequency fluctuations that occur at the modulation frequency ($f_{mod}$)  used to stabilize the 1552\,nm laser to the two-photon transition will cause a noise contribution in our demodulated error signal and hence, the frequency standard's frequency stability. 
There also needs to be a consideration of inter-modulation noise at $2nf_{mod}$~\cite{Audoin1991}. 
In our case the contribution for laser frequency noise and inter-modulation noise are the same due to the flat noise spectrum of the fiber lasers.

In order to measure the frequency noise of the 1552\,nm and 1560\,nm lasers, we optically beat each laser with another 1552\,nm and 1560\,nm laser from the same manufacturer and model.
The beat notes of the 1552\,nm and 1560\,nm lasers (at a beat frequency of approximately 10\,MHz and 60\,MHz respectively) give us the combined frequency noise of the two 1552\,nm and two 1560\,nm lasers involved in making the beat notes. 
By taking the Fourier transform of each beat note, we measured the 1560\,nm and 1552\,nm laser frequency noise spectra. 
This showed an equal frequency noise for both 1552\,nm and 1560\,nm lasers of $S(f_{mod})_{\textrm{Hz}}\,{=}\,6.5\,\text{Hz}/\sqrt{\text{Hz}}$ at our modulation frequency $f_{mod}$.
This sets an upper limit for the total frequency noise, as we measure the combination of frequency noise from two identical lasers. 
Using Eqn.\,\ref{eqn:PSDintegral} we calculate a frequency noise contribution of $3.2$Hz$/\sqrt{\tau}$ and a fractional frequency stability contribution of $8.4\,{\times}\,10^{-15}/\sqrt{\tau}$ which is the quadrature sum of both the 1552\,nm and 1560\,nm lasers. 

\subsection{Zeeman Shifts}
Each hyperfine level of the two-photon transition contains $2F{+}1$ magnetic sub-levels, where $F$ is the hyperfine level number.
At zero magnetic field these states are degenerate, but stray magnetic fields around the rubidium cell will lift this degeneracy leading to Zeeman shifts of the sub-levels.
The magnetic fields present around the rubidium cell are not high enough to spectroscopically resolve the underlying Zeeman sub-levels due to our residual Doppler broadening.
We observe a slight broadening and shift of the $5S_{1/2}\,{\rightarrow}\,5D_{5/2}$ transition, effectively coupling fluctuations in the magnetic field to instability in the frequency of the frequency standard. 
As the underlying Zeeman sub-levels are not resolved, there is no first-order dependence of the two-photon transition frequency upon magnetic field (valid for fields less than 100 mG)~\cite{Martin2018}.
A second-order shift for the 5$S_{1/2}$ $(F\,{=}\,2)$ ground state and 5$D_{5/2}$ $(F\,{=}\,4)$ excited state for the two-photon transition has previously been calculated to be 114\,Hz/G$^2$ and 50\,kHz/G$^2$ respectively~\cite{Martin2018}.

We measured the shielding factor of the $\mu$-metal shield to be at least 45-fold. 
This was measured by observing Earth’s magnetic field of 45\,$\mu$T was attenuated to below 1\,$\mu$T within the shield.
Precise measurement of the shielding factor is limited by the operational range of the magnetometer used (QuSpin QTFM sensor) which could not measure fields below 1$\mu$T. 
To estimate the magnetic field within the shield, the laboratory magnetic field environment was measured, and divided by this shielding factor.
This gives an estimate of 0.25\,mG inside the shield with a fluctuation of 3\,$\mu$G over $\tau\,{=}\,1,200$\,s.
This gives a contribution to fractional frequency stability of $3.2\,{\times}\,10^{-20}$ at $\tau\,{=}\,1200$\,s. 

\subsection{Light Shift} \label{sec:LightShift}
Light shifts (AC Stark shifts) are perturbations to the atomic energy levels due to the intensity of the lasers driving the two-photon transition~\cite{Mitroy2010, Hamilton2023}. 
A shift in the energy of the $5S_{1/2}$, $5P_{3/2}$, or $5D_{5/2}$ levels leads to a change in the frequency of the two-photon transition. 
In the two-color frequency standard configuration, contributions from both lasers are of concern.
The 780\,nm light couples predominantly to the $5\text{S}_{1/2}\,{\rightarrow}\,5\text{P}_{3/2}$ transition, frequency shifting both levels.
Whereas the 776\,nm light predominantly couples to the $5\text{P}_{3/2}\,{\rightarrow}\,5\text{D}_{5/2}$ transition, frequency shifting these two levels. 
The fractional light shift produced is proportional to the spatially averaged laser intensity $\overline{I}$ of the laser mode used to excite the transition~\cite{Martin2018,Hamilton2023}:
\begin{equation}
\label{EqnLS}
    \frac{\Delta \nu\left(\Delta_i\right)}{\nu_0} = \frac{\Delta\alpha\left(\Delta_i\right)}{2c\epsilon_0h} \overline{I}
\end{equation}
where $\Delta\nu\left(\Delta_i\right)$ is the frequency change of a transition at frequency $\nu_0$, $\Delta\alpha\left(\Delta_i\right)$ is the differential polarizability of the energy levels which incorporates laser detuning from the energy levels and polarisation of the light, $c$ is the speed of light, $\epsilon_0$ is the permittivity of free space, and $h$ is Planck's constant.

To measure the light shift sensitivity of the two-photon transition to each laser, we vary one laser power while holding the power of the other laser constant.
At an intermediate state detuning of $\Delta_i\,{=}\,1.5$\,GHz we find the light shifts induced by the 780\,nm and 776\,nm lasers to have coefficients of $-32.7\,{\pm}\,0.3$ and $2.3\,{\pm}\,0.1$\,Hz/$\mu$W respectively.
This leads to a shift per intensity of $-113\,{\pm}\,1$ and $8.0\,{\pm}\,0.4$\,kHz/(mW$\cdot$mm$^{-2}$) respectively.
This results in an absolute shift of $-19.6\,{\pm}\,0.2$ and $4.8\,{\pm}\,0.2$\,kHz respectively. 
Using these light shift coefficients together with a stability measurement of our optical powers, we are able to show that the 780\,nm light shift is the highest contribution to fractional frequency stability at longer integration times, with a contribution of $5.1\,{\times}\,10^{-15}$ at $\tau\,{=}\,1200$\,s.
By subtracting the 780\,nm light shift from the total noise, we are able to calculate a long-term stability of $4.5\,{\times}\,10^{-15}$.



We also note that as we increase our detuning from the intermediate 5$P_{3/2}$ state, we observe a reduction in magnitude of the light shift coefficients due to the reduced two-photon coupling strength, the same mechanisms that underlies Eqn.~\ref{eqn:TransRate}.
In fact, we observe that both the 776 and 780\,nm light shift coefficients decrease with increasing detuning, in a similar manner to the transition rate, as shown in Fig.~\ref{fig:Power_Detuning}, with a similar full-width half maximum (FWHM) of $\Gamma_i$ consistent with that extracted from fits shown in Fig.~\ref{fig:Power_Detuning}.
While these measurements were made very close to the 5$P_{3/2}$ intermediate state and thus are strongly influence by the hyper-fine structure of both transitions, the light shifts general behaviour is consistent with the far-detuned measurements and analysis presented in Ref.~\cite{Hamilton2023} which only considers atomic fine structure.

The light shift coefficients dependence upon intermediate state detuning was quantified by sweeping the intermediate state detuning over tens of MHz around $\Delta_i\,{=}\,1.5$\,GHz while the frequency standard is locked.
From this, we observed a light shift coefficient of 190\,Hz/MHz due to changes in the intermediate state detuning.
While the frequency standard is operational we measured the frequency stability of the 780\,nm laser as a proxy for intermediate state detuning stability.
The in-loop 780\,nm fractional frequency stability was measured to be approximately $4\,{\times}\,10^{-13}$ up to an integration time of $\tau\,{=}\,100$\,s before integrating down.
This leads to a contribution to the fractional frequency stability, from changes to the intermediate state detuning, on the order of $10^{-16}$ or less, as shown in Fig.~\ref{fig:totalSRAV}.

\subsection{Pressure Shift}
The rubidium vapor density is controlled under nominal operating conditions via heating the vapor cell to between 60$\,\degree$C and 70$\,\degree$C. 
This results in an increase of the rate of collisions between the atoms, which can be either elastic or inelastic. 
An elastic collision can cause phase changes to the atomic energy level coherence, resulting in broadening or shifting of atomic absorption lines, known as pressure broadening and pressure shifting~\cite{LEWIS1980}.
Inelastic collisions result in shortening of the two-photon excited state lifetime causing an overall broadening of the spectral line, which is also known as pressure broadening~\cite{Edwards_1973}.

By varying the temperature of the cell between approximately 60\,$\degree$C and 70\,$\degree$C, we observe a shift in the frequency of the two-photon transition to be $-460\,{\pm}\,6$\,Hz/K. 
As changing the temperature of the cell changes the rubidium vapor density, thus pressure within the vapor cell, this coefficient measures both pressure shifts and broadening.
This is in close agreement with previous measurements of $-419\,{\pm}\,15$\,Hz/K reported for a single-color two-photon frequency standard~\cite{Martin2018}. 
The discrepancy between these measurements may be due to the difference in detuning from the intermediate state~\cite{Nguyen2022}. 
The stability of our cell temperature was measured to be 0.8\,mK over $\tau\,{=}\,1200$\,s.
From this we calculate our fractional frequency stability due to pressure shifts and broadening to be $9.5\,{\times}\,10^{-16}$ at $\tau\,{=}\,1200$\,s.
Thus, this is not currently the main contributor to the frequency standard's stability, however, if light shifts can be mitigated, it would be a concern.

\subsection{Residual Amplitude Modulation}
Residual amplitude modulation (RAM) is an unintended amplitude modulation resulting from frequency modulation of the 1552\,nm laser implemented to produce a frequency discriminator.
These power fluctuations, which occur at the frequency control modulation frequency ($f_{mod}$), lead to modulations at $f_{mod}$ in both the two-photon fluorescence amplitude and generated light shifts. 
When demodulated, this produces a shift to the zero crossing lock point of the error signal the 1552\,nm laser is frequency stabilized to, see Fig.~\ref{fig:ErrorSignal}.
Thus, variations in the amount of RAM present leads to a variation in the offset of our 1552\,nm frequency lock, and thus the frequency standard's output frequency. 

To measure the impact of RAM upon the frequency stability of the standard, the out-of-loop photodetector measures the AC power, and a lock-in amplifier demodulates this signal at $f_{mod}$ to produce a measurement of the RAM amplitude.
By reducing the gain of the RAM feedback loop, we vary the amount of residual RAM remaining on the laser, then measure its effect upon the standard’s output frequency.
We determine a RAM sensitivity of 4.3\,Hz/$\mu$W.
We measured RAM fluctuations during operation to be 0.16\,$\mathrm{\mu W}$, giving a fractional frequency stability contribution of $1.8\,{\times}\,10^{-15}$ at $\tau\,{=}\,1200$\,s.

\section{Discussion}
\label{sec:Discussion}
We have presented a two-color, two-photon rubidium frequency standard, with a fractional frequency stability of $1.6\,{\times}\,10^{-13}$ at $\tau\,{=}\,1\,$s, integrating down to $8.6\,{\times}\,10^{-15}$ at $\tau\,{=}\,2400\,$s. 
A noise analysis showed the limiting factors of the frequency standard are shot noise in the short-term and 780\,nm light shift in the long-term.
By removal of the residual fluctuations of the 780\,nm laser power, thus 780\,nm light shifts, from the frequency standards output, we calculate a fractional frequency stability of $4.5\,{\times}\,10^{-15}$ at $\tau\,{=}\,2400$\,s.
This is a significant improvement on results previously presented for the two-color scheme~\cite{Perrella2019}.

The fractional frequency stability we have presented is comparable to single-color, two-photon rubidium frequency standards~\cite{Martin2018,Newman2021}, however our results have been achieved with an order of magnitude decrease in vapor number density (via decreased cell temperature), and between two~\cite{Newman2021} and ten~\cite{Martin2018} times less optical power driving the transition.
These decreases are a result of being significantly closer to the intermediate state in the two-color scheme ($\Delta_i\,{=}\,1.5$\,GHz vs $\Delta_i\,{\approx}\,1$\,THz), which increases the transition strength of the two-photon transition, see Eqn.~\ref{eqn:TransRate}.
This highlights the low power requirements of the two-color, two-photon rubidium frequency standard, making it an ideal candidate for further reduction of SWaP to produce a portable, compact frequency standard.
While these results are competitive in comparison to existing two-photon rubidium frequency standards, there are several pathways to improving short- and long-term stability of the frequency standard.

As the short-term stability is currently limited by shot noise, the stability of the frequency standard can be improved by increasing the SNR of the 420\,nm two-photon fluorescence measurement
An engineering route to improve the SNR of the two-photon transition is to collect as much of the available 420\,nm fluorescence as possible.
Increasing collection efficiency requires redesigning the physics package to utilise a more geometrically compatible rubidium cell and PMT combination which may result in an improvement in the short-term stability.

A more elegant route to improve the SNR of 420\,nm two-photon fluorescence, is to increase the amount of fluorescence produced by the atomic vapor.
To increase the two-photon transition rate, the two-color scheme has the additional parameter of intermediate state detuning to utilise, on-top of the laser power and atomic number density of the rubidium vapor that the single-color approaches use, see Eqn.~\ref{eqn:TransRate}.
Utilising this, by tuning the lasers' drive power and detuning, we have shown the potential for significantly improved fractional frequency stability of $6\,{\times}\,10^{-14}$ at $\tau\,{=}\,1\,$s.
This is a unique benefit of the two-color excitation scheme when compared to the single-color approach.

While we were able to improve the 1\,s fractional frequency stability to $6\,{\times}\,10^{-14}$ in the current configuration, we also observed a degradation to the long-term stability. 
There was a period of time between the initial measurements, shown in Fig.~\ref{fig:totalSRAV}, and the high-power measurements, shown in Figs~\ref{fig:SRAV776Power} and \ref{fig:SRAVPMTDC}. 
During this time the frequency standard developed an additional noise floor associated with alignment instability which was not present in the initial data (Fig.~\ref{fig:totalSRAV}). 
This alignment instability was verified by applying a small force to the input fiber, resulting in kHz level changes to the frequency of the frequency standard. 
We associate this with changing light shifts, due to changes in the lasers’ transverse mode and intensity in light driving the two-photon transition.
Similar effects have been observed in this two-photon transition in hollow-core fiber \cite{Perrella2013_HCF}. 
This effect cannot be negated with power and frequency control methods as described in this paper and will require the production of a new physics package, designed with this in mind.
For this reason, we were unable to obtain good long-term data for the high-power configuration, however the combination of other noise sources in conjunction with the model presented in Fig.~\ref{fig:totalSRAV} suggests we should be able to average down to $6\,{\times}\,10^{-15}$ at $\tau\,{=}\,100\,$s before encountering the combination noise sources not related to alignment instability.
This is an order of magnitude faster than has been presented in this paper, and provides a clear path forward for research on this frequency standard.

At longer integration times, the stability of the frequency standard is limited by 780\,nm light shifts.
This is particularly evident at low intermediate state detunings or higher optical powers, both of which lead to an increase in the magnitude of the light shifts.
To control these light shifts, future work should focus on improvements in the feedback loops controlling the optical power of the lasers, specifically the RAM suppression, AC power stabilization of the 780\,nm light, and DC power stabilization of the 776\,nm light.
Another possible method to decreasing the long-term contribution of light shift may be auto-balanced Ramsey spectroscopy~\cite{Sanner2018} or other spectroscopic methods~\cite{Yudin2018}. 
Reducing the effects of light shifts will benefit the stability of the frequency standard at both near and far intermediate state detunings. 
This could remove light shifts as the dominant contribution above 1000\,s integration time as shown in Fig.~\ref{fig:totalSRAV} with pressure shifts being the next largest contribution.

An additional improvement to the long-term stability of the standard would be to improve the temperature control of the rubidium cell.
This could be achieved through improving passive isolation to the environment and improving the temperature control loop. 
Alternatively, recent theoretical work has shown that, to first order, pressure shifts may be able to be cancelled by a frequency pulling effect via tuning of the intermediate state detuning and cell temperature~\cite{Nguyen2022}.  
This would greatly benefit deployable frequency standards in uncontrolled environments. 
Tailoring of detuning and temperatures as discussed in Ref~\cite{Nguyen2022} could allow higher operational temperatures without the deleterious increase in pressure shifts which increase with temperature. 
Increasing the temperature would improve SNR leading to improved shot noise and short-term stabilities. 
This was not explored here as in the current design many components are approaching their maximum temperature rating. 
A redesign may enable operation at higher temperatures in the future.

\section{Conclusion} \label{sec:conclusion}
We have presented a two-color, two-photon, rubidium frequency standard based on the $5S_{1/2}\,{\rightarrow}\,5D_{5/2}$ transition.
We show a short-term stability of $1.6\,{\times}\,10^{-13}$ at $\tau\,{=}\,1$\,s, and a long-term stability of $8.6\,{\times}\,10^{-15}$ at $\tau\,{=}\,2400$\,s.
Through the subtraction of residual 780\,nm light shifts, we calculate a stability of $4.5\,{\times}\,10^{-15}$ at $\tau\,{=}\,2400$\,s.
By utilising the two-color excitation, we are able to achieve these stabilities with ten-fold less optical power and ten-fold lower rubidium vapor density, compared to previous rubidium single-color two-photon frequency standard demonstrations.
We demonstrate the advantage of the two-color approach by optimising the detuning from the $5P_{3/2}$ intermediate state and optical powers driving the two-photon transition, allowing us to demonstrate a stability of $6\,{\times}\,10^{-14}$ at $\tau\,{=}\,1$\,s.
The noise sources of the frequency standard have been characterized, showing the short-term stability is limited by shot noise, while the long-term stability is limited by light shifts.
An integrated optical frequency comb makes the system fully self-referenced, removing the need for any external frequency references, and down-converts the frequency standard's stable optical output to the microwave domain, compatible with PNT technologies. 
This work represents an advance towards a low size, weight, and power frequency standard based on this two-color excitation method.

\acknowledgments  
This work has been supported by the SmartSat CRC, whose activities are funded by the Australian Government’s CRC Program. 
We acknowledge the support of the South Australian Government via the Industrial Doctoral Training Centre program.
This research was conducted by the Australian Research Council Centre of Excellence in Optical Microcombs for Breakthrough Science (project number CE230100006) and funded by the Australian Government.
We acknowledge support from the U.S. AFOSR AOARD FA2386-19-1-4054 and FA2386-20-1-4032 grants.
This research is supported by the Commonwealth of Australia Defence Science and Technology Group.
We are grateful for the invaluable input from Ashby Hilton, Rachel Offer, Montana Nelligen and the rest of the Precision Measurement Group at the Institute for Photonics and Advanced Sensing (IPAS) at the University of Adelaide.
We thank Prof. Kishan Dholakia for critical reading of the article.
We thank the Optofab node of the Australian National Fabrication Facility (ANFF) which utilize Commonwealth and South Australia State Government funding. 
The authors thank Evan Johnson, Alastair Dowler, and Lijesh Thomas and the rest of the Optofab team for their technical support. 
The work presented is covered by patent US 10353270 B2.


\begin{thebibliography}{58}%
\makeatletter
\providecommand \@ifxundefined [1]{%
 \@ifx{#1\undefined}
}%
\providecommand \@ifnum [1]{%
 \ifnum #1\expandafter \@firstoftwo
 \else \expandafter \@secondoftwo
 \fi
}%
\providecommand \@ifx [1]{%
 \ifx #1\expandafter \@firstoftwo
 \else \expandafter \@secondoftwo
 \fi
}%
\providecommand \natexlab [1]{#1}%
\providecommand \enquote  [1]{``#1''}%
\providecommand \bibnamefont  [1]{#1}%
\providecommand \bibfnamefont [1]{#1}%
\providecommand \citenamefont [1]{#1}%
\providecommand \href@noop [0]{\@secondoftwo}%
\providecommand \href [0]{\begingroup \@sanitize@url \@href}%
\providecommand \@href[1]{\@@startlink{#1}\@@href}%
\providecommand \@@href[1]{\endgroup#1\@@endlink}%
\providecommand \@sanitize@url [0]{\catcode `\\12\catcode `\$12\catcode
  `\&12\catcode `\#12\catcode `\^12\catcode `\_12\catcode `\%12\relax}%
\providecommand \@@startlink[1]{}%
\providecommand \@@endlink[0]{}%
\providecommand \url  [0]{\begingroup\@sanitize@url \@url }%
\providecommand \@url [1]{\endgroup\@href {#1}{\urlprefix }}%
\providecommand \urlprefix  [0]{URL }%
\providecommand \Eprint [0]{\href }%
\providecommand \doibase [0]{https://doi.org/}%
\providecommand \selectlanguage [0]{\@gobble}%
\providecommand \bibinfo  [0]{\@secondoftwo}%
\providecommand \bibfield  [0]{\@secondoftwo}%
\providecommand \translation [1]{[#1]}%
\providecommand \BibitemOpen [0]{}%
\providecommand \bibitemStop [0]{}%
\providecommand \bibitemNoStop [0]{.\EOS\space}%
\providecommand \EOS [0]{\spacefactor3000\relax}%
\providecommand \BibitemShut  [1]{\csname bibitem#1\endcsname}%
\let\auto@bib@innerbib\@empty
\bibitem [{\citenamefont {Ramsey}(2005)}]{Ramsey_2005}%
  \BibitemOpen
  \bibfield  {author} {\bibinfo {author} {\bibfnamefont {N.~F.}\ \bibnamefont
  {Ramsey}},\ }\bibfield  {title} {\bibinfo {title} {History of early atomic
  clocks},\ }\href {https://doi.org/10.1088/0026-1394/42/3/S01} {\bibfield
  {journal} {\bibinfo  {journal} {Metrologia}\ }\textbf {\bibinfo {volume}
  {42}},\ \bibinfo {pages} {S1} (\bibinfo {year} {2005})}\BibitemShut {NoStop}%
\bibitem [{\citenamefont {Essen}\ and\ \citenamefont
  {Parry}(1955)}]{essen1955atomic}%
  \BibitemOpen
  \bibfield  {author} {\bibinfo {author} {\bibfnamefont {L.}~\bibnamefont
  {Essen}}\ and\ \bibinfo {author} {\bibfnamefont {J.~V.}\ \bibnamefont
  {Parry}},\ }\bibfield  {title} {\bibinfo {title} {An atomic standard of
  frequency and time interval: a caesium resonator},\ }\href@noop {} {\bibfield
   {journal} {\bibinfo  {journal} {Nature}\ }\textbf {\bibinfo {volume}
  {176}},\ \bibinfo {pages} {280} (\bibinfo {year} {1955})}\BibitemShut
  {NoStop}%
\bibitem [{\citenamefont {Lombardi}(2017)}]{lombardi2017historical}%
  \BibitemOpen
  \bibfield  {author} {\bibinfo {author} {\bibfnamefont {M.~A.}\ \bibnamefont
  {Lombardi}},\ }\bibfield  {title} {\bibinfo {title} {A historical review of
  us contributions to the atomic redefinition of the si second in 1967},\
  }\href@noop {} {\bibfield  {journal} {\bibinfo  {journal} {Journal of
  Research of the National Institute of Standards and Technology}\ }\textbf
  {\bibinfo {volume} {122}},\ \bibinfo {pages} {1} (\bibinfo {year}
  {2017})}\BibitemShut {NoStop}%
\bibitem [{\citenamefont {Mallette}\ \emph {et~al.}(2010)\citenamefont
  {Mallette}, \citenamefont {White},\ and\ \citenamefont {Rochat}}]{Mallette}%
  \BibitemOpen
  \bibfield  {author} {\bibinfo {author} {\bibfnamefont {L.~A.}\ \bibnamefont
  {Mallette}}, \bibinfo {author} {\bibfnamefont {J.}~\bibnamefont {White}},\
  and\ \bibinfo {author} {\bibfnamefont {P.}~\bibnamefont {Rochat}},\
  }\bibfield  {title} {\bibinfo {title} {Space qualified frequency sources
  (clocks) for current and future gnss applications},\ }in\ \href
  {https://doi.org/10.1109/PLANS.2010.5507225} {\emph {\bibinfo {booktitle}
  {IEEE/ION Position, Location and Navigation Symposium}}}\ (\bibinfo {year}
  {2010})\ pp.\ \bibinfo {pages} {903--908}\BibitemShut {NoStop}%
\bibitem [{\citenamefont {Batori}\ \emph {et~al.}(2020)\citenamefont {Batori},
  \citenamefont {Almat}, \citenamefont {Affolderbach},\ and\ \citenamefont
  {Mileti}}]{Batori2020}%
  \BibitemOpen
  \bibfield  {author} {\bibinfo {author} {\bibfnamefont {E.}~\bibnamefont
  {Batori}}, \bibinfo {author} {\bibfnamefont {N.}~\bibnamefont {Almat}},
  \bibinfo {author} {\bibfnamefont {C.}~\bibnamefont {Affolderbach}},\ and\
  \bibinfo {author} {\bibfnamefont {G.}~\bibnamefont {Mileti}},\ }\bibfield
  {title} {\bibinfo {title} {{GNSS-grade space atomic frequency standards:
  Current status and ongoing developments}},\ }\bibfield  {journal} {\bibinfo
  {journal} {Advances in Space Research}\ }\href
  {https://doi.org/10.1016/j.asr.2020.09.012} {10.1016/j.asr.2020.09.012}
  (\bibinfo {year} {2020})\BibitemShut {NoStop}%
\bibitem [{\citenamefont {Jaduszliwer}\ and\ \citenamefont
  {Camparo}(2021)}]{Jaduszliwer2021}%
  \BibitemOpen
  \bibfield  {author} {\bibinfo {author} {\bibfnamefont {B.}~\bibnamefont
  {Jaduszliwer}}\ and\ \bibinfo {author} {\bibfnamefont {J.}~\bibnamefont
  {Camparo}},\ }\bibfield  {title} {\bibinfo {title} {{Past, present and future
  of atomic clocks for GNSS}},\ }\href
  {https://doi.org/10.1007/s10291-020-01059-x} {\bibfield  {journal} {\bibinfo
  {journal} {GPS Solutions}\ }\textbf {\bibinfo {volume} {25}},\ \bibinfo
  {pages} {1} (\bibinfo {year} {2021})}\BibitemShut {NoStop}%
\bibitem [{\citenamefont {Schuldt}\ \emph {et~al.}(2020)\citenamefont
  {Schuldt}, \citenamefont {Gohlke}, \citenamefont {Oswald}, \citenamefont
  {Sanjuan}, \citenamefont {Wegehaupt}, \citenamefont {Blomberg}, \citenamefont
  {Wust}, \citenamefont {Blumel}, \citenamefont {Gualani}, \citenamefont
  {Abich},\ and\ \citenamefont {Braxmaier}}]{Schuldt2020}%
  \BibitemOpen
  \bibfield  {author} {\bibinfo {author} {\bibfnamefont {T.}~\bibnamefont
  {Schuldt}}, \bibinfo {author} {\bibfnamefont {M.}~\bibnamefont {Gohlke}},
  \bibinfo {author} {\bibfnamefont {M.}~\bibnamefont {Oswald}}, \bibinfo
  {author} {\bibfnamefont {J.}~\bibnamefont {Sanjuan}}, \bibinfo {author}
  {\bibfnamefont {T.}~\bibnamefont {Wegehaupt}}, \bibinfo {author}
  {\bibfnamefont {T.}~\bibnamefont {Blomberg}}, \bibinfo {author}
  {\bibfnamefont {J.}~\bibnamefont {Wust}}, \bibinfo {author} {\bibfnamefont
  {L.}~\bibnamefont {Blumel}}, \bibinfo {author} {\bibfnamefont
  {V.}~\bibnamefont {Gualani}}, \bibinfo {author} {\bibfnamefont
  {K.}~\bibnamefont {Abich}},\ and\ \bibinfo {author} {\bibfnamefont
  {C.}~\bibnamefont {Braxmaier}},\ }\bibfield  {title} {\bibinfo {title}
  {{Optical clock technologies enabling advanced GNSS}},\ }\href
  {https://doi.org/10.23919/ENC48637.2020.9317452} {\bibfield  {journal}
  {\bibinfo  {journal} {2020 European Navigation Conference, ENC 2020}\ ,\
  \bibinfo {pages} {2}} (\bibinfo {year} {2020})}\BibitemShut {NoStop}%
\bibitem [{\citenamefont {Hollberg}(2020)}]{Hollberg2021}%
  \BibitemOpen
  \bibfield  {author} {\bibinfo {author} {\bibfnamefont {L.}~\bibnamefont
  {Hollberg}},\ }\bibinfo {title} {Atomic clocks for gnss},\ in\ \href
  {https://doi.org/https://doi.org/10.1002/9781119458555.ch47} {\emph {\bibinfo
  {booktitle} {Atomic Clocks for GNSS. In Position, Navigation, and Timing
  Technologies in the 21st Century}}}\ (\bibinfo  {publisher} {John Wiley \&
  Sons, Ltd},\ \bibinfo {year} {2020})\ Chap.~\bibinfo {chapter} {47}, pp.\
  \bibinfo {pages} {1497--1519}\BibitemShut {NoStop}%
\bibitem [{\citenamefont {Riley}(2019)}]{riley2019history}%
  \BibitemOpen
  \bibfield  {author} {\bibinfo {author} {\bibfnamefont {W.~J.}\ \bibnamefont
  {Riley}},\ }\bibfield  {title} {\bibinfo {title} {A history of the rubidium
  frequency standard},\ }\href@noop {} {\bibfield  {journal} {\bibinfo
  {journal} {IEEE UFFC-S History}\ ,\ \bibinfo {pages} {2}} (\bibinfo {year}
  {2019})}\BibitemShut {NoStop}%
\bibitem [{\citenamefont {Gill}(2005)}]{gill2005optical}%
  \BibitemOpen
  \bibfield  {author} {\bibinfo {author} {\bibfnamefont {P.}~\bibnamefont
  {Gill}},\ }\bibfield  {title} {\bibinfo {title} {Optical frequency
  standards},\ }\href@noop {} {\bibfield  {journal} {\bibinfo  {journal}
  {Metrologia}\ }\textbf {\bibinfo {volume} {42}},\ \bibinfo {pages} {S125}
  (\bibinfo {year} {2005})}\BibitemShut {NoStop}%
\bibitem [{\citenamefont {Burt}\ \emph {et~al.}(2021)\citenamefont {Burt},
  \citenamefont {Prestage}, \citenamefont {Tjoelker}, \citenamefont {Enzer},
  \citenamefont {Kuang}, \citenamefont {Murphy}, \citenamefont {Robison},
  \citenamefont {Seubert}, \citenamefont {Wang},\ and\ \citenamefont
  {Ely}}]{Burt2021}%
  \BibitemOpen
  \bibfield  {author} {\bibinfo {author} {\bibfnamefont {E.~A.}\ \bibnamefont
  {Burt}}, \bibinfo {author} {\bibfnamefont {J.~D.}\ \bibnamefont {Prestage}},
  \bibinfo {author} {\bibfnamefont {R.~L.}\ \bibnamefont {Tjoelker}}, \bibinfo
  {author} {\bibfnamefont {D.~G.}\ \bibnamefont {Enzer}}, \bibinfo {author}
  {\bibfnamefont {D.}~\bibnamefont {Kuang}}, \bibinfo {author} {\bibfnamefont
  {D.~W.}\ \bibnamefont {Murphy}}, \bibinfo {author} {\bibfnamefont {D.~E.}\
  \bibnamefont {Robison}}, \bibinfo {author} {\bibfnamefont {J.~M.}\
  \bibnamefont {Seubert}}, \bibinfo {author} {\bibfnamefont {R.~T.}\
  \bibnamefont {Wang}},\ and\ \bibinfo {author} {\bibfnamefont {T.~A.}\
  \bibnamefont {Ely}},\ }\bibfield  {title} {\bibinfo {title} {{Demonstration
  of a trapped-ion atomic clock in space}},\ }\href
  {https://doi.org/10.1038/s41586-021-03571-7} {\bibfield  {journal} {\bibinfo
  {journal} {Nature}\ }\textbf {\bibinfo {volume} {595}},\ \bibinfo {pages}
  {43} (\bibinfo {year} {2021})}\BibitemShut {NoStop}%
\bibitem [{\citenamefont {Hao}\ \emph {et~al.}(2024)\citenamefont {Hao},
  \citenamefont {Yang}, \citenamefont {Ruan}, \citenamefont {Yun},\ and\
  \citenamefont {Zhang}}]{Hao2024}%
  \BibitemOpen
  \bibfield  {author} {\bibinfo {author} {\bibfnamefont {Q.}~\bibnamefont
  {Hao}}, \bibinfo {author} {\bibfnamefont {S.}~\bibnamefont {Yang}}, \bibinfo
  {author} {\bibfnamefont {J.}~\bibnamefont {Ruan}}, \bibinfo {author}
  {\bibfnamefont {P.}~\bibnamefont {Yun}},\ and\ \bibinfo {author}
  {\bibfnamefont {S.}~\bibnamefont {Zhang}},\ }\bibfield  {title} {\bibinfo
  {title} {Integrated pulsed optically pumped rb atomic clock with frequency
  stability of ${10}^{\ensuremath{-}15}$},\ }\href
  {https://doi.org/10.1103/PhysRevApplied.21.024003} {\bibfield  {journal}
  {\bibinfo  {journal} {Phys. Rev. Appl.}\ }\textbf {\bibinfo {volume} {21}},\
  \bibinfo {pages} {024003} (\bibinfo {year} {2024})}\BibitemShut {NoStop}%
\bibitem [{\citenamefont {Ye}\ \emph {et~al.}(2001)\citenamefont {Ye},
  \citenamefont {Ma},\ and\ \citenamefont {Hall}}]{ye2001}%
  \BibitemOpen
  \bibfield  {author} {\bibinfo {author} {\bibfnamefont {J.}~\bibnamefont
  {Ye}}, \bibinfo {author} {\bibfnamefont {L.~S.}\ \bibnamefont {Ma}},\ and\
  \bibinfo {author} {\bibfnamefont {J.~L.}\ \bibnamefont {Hall}},\ }\bibfield
  {title} {\bibinfo {title} {Molecular iodine clock},\ }\href
  {https://doi.org/10.1103/PhysRevLett.87.270801} {\bibfield  {journal}
  {\bibinfo  {journal} {Phys. Rev. Lett.}\ }\textbf {\bibinfo {volume} {87}},\
  \bibinfo {pages} {270801} (\bibinfo {year} {2001})}\BibitemShut {NoStop}%
\bibitem [{\citenamefont {Diddams}\ \emph {et~al.}(2001)\citenamefont
  {Diddams}, \citenamefont {Udem}, \citenamefont {Bergquist}, \citenamefont
  {Curtis}, \citenamefont {Drullinger}, \citenamefont {Hollberg}, \citenamefont
  {Itano}, \citenamefont {Lee}, \citenamefont {Oates}, \citenamefont {Vogel},\
  and\ \citenamefont {Wineland}}]{diddams2001}%
  \BibitemOpen
  \bibfield  {author} {\bibinfo {author} {\bibfnamefont {S.~A.}\ \bibnamefont
  {Diddams}}, \bibinfo {author} {\bibfnamefont {T.}~\bibnamefont {Udem}},
  \bibinfo {author} {\bibfnamefont {J.~C.}\ \bibnamefont {Bergquist}}, \bibinfo
  {author} {\bibfnamefont {E.~A.}\ \bibnamefont {Curtis}}, \bibinfo {author}
  {\bibfnamefont {R.~E.}\ \bibnamefont {Drullinger}}, \bibinfo {author}
  {\bibfnamefont {L.}~\bibnamefont {Hollberg}}, \bibinfo {author}
  {\bibfnamefont {W.~M.}\ \bibnamefont {Itano}}, \bibinfo {author}
  {\bibfnamefont {W.~D.}\ \bibnamefont {Lee}}, \bibinfo {author} {\bibfnamefont
  {C.~W.}\ \bibnamefont {Oates}}, \bibinfo {author} {\bibfnamefont {K.~R.}\
  \bibnamefont {Vogel}},\ and\ \bibinfo {author} {\bibfnamefont {D.~J.}\
  \bibnamefont {Wineland}},\ }\bibfield  {title} {\bibinfo {title} {An optical
  clock based on a single trapped $^{199}\text{Hg}^+$ ion},\ }\href
  {https://doi.org/10.1126/science.1061171} {\bibfield  {journal} {\bibinfo
  {journal} {Science}\ }\textbf {\bibinfo {volume} {293}},\ \bibinfo {pages}
  {825} (\bibinfo {year} {2001})}\BibitemShut {NoStop}%
\bibitem [{\citenamefont {Jones}\ \emph {et~al.}(2000)\citenamefont {Jones},
  \citenamefont {Diddams}, \citenamefont {Ranka}, \citenamefont {Stentz},
  \citenamefont {Windeler}, \citenamefont {Hall},\ and\ \citenamefont
  {Cundiff}}]{Jones2000}%
  \BibitemOpen
  \bibfield  {author} {\bibinfo {author} {\bibfnamefont {D.~J.}\ \bibnamefont
  {Jones}}, \bibinfo {author} {\bibfnamefont {S.~A.}\ \bibnamefont {Diddams}},
  \bibinfo {author} {\bibfnamefont {J.~K.}\ \bibnamefont {Ranka}}, \bibinfo
  {author} {\bibfnamefont {A.}~\bibnamefont {Stentz}}, \bibinfo {author}
  {\bibfnamefont {R.~S.}\ \bibnamefont {Windeler}}, \bibinfo {author}
  {\bibfnamefont {J.~L.}\ \bibnamefont {Hall}},\ and\ \bibinfo {author}
  {\bibfnamefont {S.~T.}\ \bibnamefont {Cundiff}},\ }\bibfield  {title}
  {\bibinfo {title} {Carrier-envelope phase control of femtosecond mode-locked
  lasers and direct optical frequency synthesis},\ }\href
  {https://doi.org/10.1126/science.288.5466.635} {\bibfield  {journal}
  {\bibinfo  {journal} {Science}\ }\textbf {\bibinfo {volume} {288}},\ \bibinfo
  {pages} {635} (\bibinfo {year} {2000})}\BibitemShut {NoStop}%
\bibitem [{\citenamefont {Hutson}\ \emph {et~al.}(2024)\citenamefont {Hutson},
  \citenamefont {Milner}, \citenamefont {Yan}, \citenamefont {Ye},\ and\
  \citenamefont {Sanner}}]{Hutson2024}%
  \BibitemOpen
  \bibfield  {author} {\bibinfo {author} {\bibfnamefont {R.~B.}\ \bibnamefont
  {Hutson}}, \bibinfo {author} {\bibfnamefont {W.~R.}\ \bibnamefont {Milner}},
  \bibinfo {author} {\bibfnamefont {L.}~\bibnamefont {Yan}}, \bibinfo {author}
  {\bibfnamefont {J.}~\bibnamefont {Ye}},\ and\ \bibinfo {author}
  {\bibfnamefont {C.}~\bibnamefont {Sanner}},\ }\bibfield  {title} {\bibinfo
  {title} {Observation of millihertz-level cooperative lamb shifts in an
  optical atomic clock},\ }\href {https://doi.org/10.1126/science.adh4477}
  {\bibfield  {journal} {\bibinfo  {journal} {Science}\ }\textbf {\bibinfo
  {volume} {383}},\ \bibinfo {pages} {384} (\bibinfo {year}
  {2024})}\BibitemShut {NoStop}%
\bibitem [{\citenamefont {Brewer}\ \emph {et~al.}(2019)\citenamefont {Brewer},
  \citenamefont {Chen}, \citenamefont {Hankin}, \citenamefont {Clements},
  \citenamefont {Chou}, \citenamefont {Wineland}, \citenamefont {Hume},\ and\
  \citenamefont {Leibrandt}}]{Brewer2019}%
  \BibitemOpen
  \bibfield  {author} {\bibinfo {author} {\bibfnamefont {S.~M.}\ \bibnamefont
  {Brewer}}, \bibinfo {author} {\bibfnamefont {J.-S.}\ \bibnamefont {Chen}},
  \bibinfo {author} {\bibfnamefont {A.~M.}\ \bibnamefont {Hankin}}, \bibinfo
  {author} {\bibfnamefont {E.~R.}\ \bibnamefont {Clements}}, \bibinfo {author}
  {\bibfnamefont {C.~W.}\ \bibnamefont {Chou}}, \bibinfo {author}
  {\bibfnamefont {D.~J.}\ \bibnamefont {Wineland}}, \bibinfo {author}
  {\bibfnamefont {D.~B.}\ \bibnamefont {Hume}},\ and\ \bibinfo {author}
  {\bibfnamefont {D.~R.}\ \bibnamefont {Leibrandt}},\ }\bibfield  {title}
  {\bibinfo {title} {$^{27}{\mathrm{al}}^{+}$ quantum-logic clock with a
  systematic uncertainty below ${10}^{\ensuremath{-}18}$},\ }\href
  {https://doi.org/10.1103/PhysRevLett.123.033201} {\bibfield  {journal}
  {\bibinfo  {journal} {Phys. Rev. Lett.}\ }\textbf {\bibinfo {volume} {123}},\
  \bibinfo {pages} {033201} (\bibinfo {year} {2019})}\BibitemShut {NoStop}%
\bibitem [{\citenamefont {Safronova}\ \emph {et~al.}(2018)\citenamefont
  {Safronova}, \citenamefont {Budker}, \citenamefont {DeMille}, \citenamefont
  {Kimball}, \citenamefont {Derevianko},\ and\ \citenamefont
  {Clark}}]{Safronova2018}%
  \BibitemOpen
  \bibfield  {author} {\bibinfo {author} {\bibfnamefont {M.~S.}\ \bibnamefont
  {Safronova}}, \bibinfo {author} {\bibfnamefont {D.}~\bibnamefont {Budker}},
  \bibinfo {author} {\bibfnamefont {D.}~\bibnamefont {DeMille}}, \bibinfo
  {author} {\bibfnamefont {D.~F.~J.}\ \bibnamefont {Kimball}}, \bibinfo
  {author} {\bibfnamefont {A.}~\bibnamefont {Derevianko}},\ and\ \bibinfo
  {author} {\bibfnamefont {C.~W.}\ \bibnamefont {Clark}},\ }\bibfield  {title}
  {\bibinfo {title} {Search for new physics with atoms and molecules},\ }\href
  {https://doi.org/10.1103/RevModPhys.90.025008} {\bibfield  {journal}
  {\bibinfo  {journal} {Rev. Mod. Phys.}\ }\textbf {\bibinfo {volume} {90}},\
  \bibinfo {pages} {025008} (\bibinfo {year} {2018})}\BibitemShut {NoStop}%
\bibitem [{\citenamefont {Katori}(2011)}]{Katori2011}%
  \BibitemOpen
  \bibfield  {author} {\bibinfo {author} {\bibfnamefont {H.}~\bibnamefont
  {Katori}},\ }\bibfield  {title} {\bibinfo {title} {Optical lattice clocks and
  quantum metrology},\ }\href {https://doi.org/10.1038/nphoton.2011.45}
  {\bibfield  {journal} {\bibinfo  {journal} {Nature Photonics}\ }\textbf
  {\bibinfo {volume} {5}} (\bibinfo {year} {2011})}\BibitemShut {NoStop}%
\bibitem [{\citenamefont {Bize}\ \emph {et~al.}(2003)\citenamefont {Bize},
  \citenamefont {Diddams}, \citenamefont {Tanaka}, \citenamefont {Tanner},
  \citenamefont {Oskay}, \citenamefont {Drullinger}, \citenamefont {Parker},
  \citenamefont {Heavner}, \citenamefont {Jefferts}, \citenamefont {Hollberg},
  \citenamefont {Itano},\ and\ \citenamefont {Bergquist}}]{Bize2003}%
  \BibitemOpen
  \bibfield  {author} {\bibinfo {author} {\bibfnamefont {S.}~\bibnamefont
  {Bize}}, \bibinfo {author} {\bibfnamefont {S.~A.}\ \bibnamefont {Diddams}},
  \bibinfo {author} {\bibfnamefont {U.}~\bibnamefont {Tanaka}}, \bibinfo
  {author} {\bibfnamefont {C.~E.}\ \bibnamefont {Tanner}}, \bibinfo {author}
  {\bibfnamefont {W.~H.}\ \bibnamefont {Oskay}}, \bibinfo {author}
  {\bibfnamefont {R.~E.}\ \bibnamefont {Drullinger}}, \bibinfo {author}
  {\bibfnamefont {T.~E.}\ \bibnamefont {Parker}}, \bibinfo {author}
  {\bibfnamefont {T.~P.}\ \bibnamefont {Heavner}}, \bibinfo {author}
  {\bibfnamefont {S.~R.}\ \bibnamefont {Jefferts}}, \bibinfo {author}
  {\bibfnamefont {L.}~\bibnamefont {Hollberg}}, \bibinfo {author}
  {\bibfnamefont {W.~M.}\ \bibnamefont {Itano}},\ and\ \bibinfo {author}
  {\bibfnamefont {J.~C.}\ \bibnamefont {Bergquist}},\ }\bibfield  {title}
  {\bibinfo {title} {{Testing the Stability of Fundamental Constants with the
  $^{199}\text{Hg}^+$ Single-Ion Optical Clock}},\ }\href
  {https://doi.org/10.1103/PhysRevLett.90.150802} {\bibfield  {journal}
  {\bibinfo  {journal} {Physical Review Letters}\ }\textbf {\bibinfo {volume}
  {90}},\ \bibinfo {pages} {4} (\bibinfo {year} {2003})}\BibitemShut {NoStop}%
\bibitem [{\citenamefont {Reinhardt}\ \emph {et~al.}(2007)\citenamefont
  {Reinhardt}, \citenamefont {Saathoff}, \citenamefont {Buhr}, \citenamefont
  {Carlson}, \citenamefont {Wolf}, \citenamefont {Schwalm}, \citenamefont
  {Karpuk}, \citenamefont {Novotny}, \citenamefont {Huber}, \citenamefont
  {Zimmermann}, \citenamefont {Holzwarth}, \citenamefont {Udem}, \citenamefont
  {H{\"{a}}nsch},\ and\ \citenamefont {Gwinner}}]{Reinhardt2007}%
  \BibitemOpen
  \bibfield  {author} {\bibinfo {author} {\bibfnamefont {S.}~\bibnamefont
  {Reinhardt}}, \bibinfo {author} {\bibfnamefont {G.}~\bibnamefont {Saathoff}},
  \bibinfo {author} {\bibfnamefont {H.}~\bibnamefont {Buhr}}, \bibinfo {author}
  {\bibfnamefont {L.~A.}\ \bibnamefont {Carlson}}, \bibinfo {author}
  {\bibfnamefont {A.}~\bibnamefont {Wolf}}, \bibinfo {author} {\bibfnamefont
  {D.}~\bibnamefont {Schwalm}}, \bibinfo {author} {\bibfnamefont
  {S.}~\bibnamefont {Karpuk}}, \bibinfo {author} {\bibfnamefont
  {C.}~\bibnamefont {Novotny}}, \bibinfo {author} {\bibfnamefont
  {G.}~\bibnamefont {Huber}}, \bibinfo {author} {\bibfnamefont
  {M.}~\bibnamefont {Zimmermann}}, \bibinfo {author} {\bibfnamefont
  {R.}~\bibnamefont {Holzwarth}}, \bibinfo {author} {\bibfnamefont
  {T.}~\bibnamefont {Udem}}, \bibinfo {author} {\bibfnamefont {T.~W.}\
  \bibnamefont {H{\"{a}}nsch}},\ and\ \bibinfo {author} {\bibfnamefont
  {G.}~\bibnamefont {Gwinner}},\ }\bibfield  {title} {\bibinfo {title} {{Test
  of relativistic time dilation with fast optical atomic clocks at different
  velocities}},\ }\href {https://doi.org/10.1038/nphys778} {\bibfield
  {journal} {\bibinfo  {journal} {Nature Physics}\ }\textbf {\bibinfo {volume}
  {3}},\ \bibinfo {pages} {861} (\bibinfo {year} {2007})}\BibitemShut {NoStop}%
\bibitem [{\citenamefont {Riehle}(2015)}]{Riehle2015}%
  \BibitemOpen
  \bibfield  {author} {\bibinfo {author} {\bibfnamefont {F.}~\bibnamefont
  {Riehle}},\ }\bibfield  {title} {\bibinfo {title} {Towards a re-definition of
  the second based on optical atomic clocks},\ }\href
  {https://doi.org/10.1016/j.crhy.2015.03.012} {\bibfield  {journal} {\bibinfo
  {journal} {Comptes Rendus Physique}\ }\textbf {\bibinfo {volume} {16}}
  (\bibinfo {year} {2015})}\BibitemShut {NoStop}%
\bibitem [{\citenamefont {Bregolin}\ \emph {et~al.}(2017)\citenamefont
  {Bregolin}, \citenamefont {Milani}, \citenamefont {Pizzocaro}, \citenamefont
  {Rauf}, \citenamefont {Thoumany}, \citenamefont {Levi},\ and\ \citenamefont
  {Calonico}}]{bregolin2017optical}%
  \BibitemOpen
  \bibfield  {author} {\bibinfo {author} {\bibfnamefont {F.}~\bibnamefont
  {Bregolin}}, \bibinfo {author} {\bibfnamefont {G.}~\bibnamefont {Milani}},
  \bibinfo {author} {\bibfnamefont {M.}~\bibnamefont {Pizzocaro}}, \bibinfo
  {author} {\bibfnamefont {B.}~\bibnamefont {Rauf}}, \bibinfo {author}
  {\bibfnamefont {P.}~\bibnamefont {Thoumany}}, \bibinfo {author}
  {\bibfnamefont {F.}~\bibnamefont {Levi}},\ and\ \bibinfo {author}
  {\bibfnamefont {D.}~\bibnamefont {Calonico}},\ }\bibfield  {title} {\bibinfo
  {title} {Optical lattice clocks towards the redefinition of the second},\
  }in\ \href@noop {} {\emph {\bibinfo {booktitle} {Journal of Physics:
  Conference Series}}},\ Vol.\ \bibinfo {volume} {841}\ (\bibinfo
  {organization} {IOP Publishing},\ \bibinfo {year} {2017})\ p.\ \bibinfo
  {pages} {012015}\BibitemShut {NoStop}%
\bibitem [{\citenamefont {Poli}\ \emph {et~al.}(2014)\citenamefont {Poli},
  \citenamefont {Schioppo}, \citenamefont {Vogt}, \citenamefont {Falke},
  \citenamefont {Sterr}, \citenamefont {Lisdat},\ and\ \citenamefont
  {Tino}}]{Poli2014}%
  \BibitemOpen
  \bibfield  {author} {\bibinfo {author} {\bibfnamefont {N.}~\bibnamefont
  {Poli}}, \bibinfo {author} {\bibfnamefont {M.}~\bibnamefont {Schioppo}},
  \bibinfo {author} {\bibfnamefont {S.}~\bibnamefont {Vogt}}, \bibinfo {author}
  {\bibfnamefont {S.}~\bibnamefont {Falke}}, \bibinfo {author} {\bibfnamefont
  {U.}~\bibnamefont {Sterr}}, \bibinfo {author} {\bibfnamefont
  {C.}~\bibnamefont {Lisdat}},\ and\ \bibinfo {author} {\bibfnamefont {G.~M.}\
  \bibnamefont {Tino}},\ }\bibfield  {title} {\bibinfo {title} {{A
  transportable strontium optical lattice clock}},\ }\href
  {https://doi.org/10.1007/s00340-014-5932-9} {\bibfield  {journal} {\bibinfo
  {journal} {Applied Physics B}\ }\textbf {\bibinfo {volume} {117}},\ \bibinfo
  {pages} {1107} (\bibinfo {year} {2014})}\BibitemShut {NoStop}%
\bibitem [{\citenamefont {Koller}\ \emph {et~al.}(2017)\citenamefont {Koller},
  \citenamefont {Grotti}, \citenamefont {Vogt}, \citenamefont {Al-Masoudi},
  \citenamefont {D\"orscher}, \citenamefont {H\"afner}, \citenamefont {Sterr},\
  and\ \citenamefont {Lisdat}}]{Koller2017}%
  \BibitemOpen
  \bibfield  {author} {\bibinfo {author} {\bibfnamefont {S.~B.}\ \bibnamefont
  {Koller}}, \bibinfo {author} {\bibfnamefont {J.}~\bibnamefont {Grotti}},
  \bibinfo {author} {\bibfnamefont {S.}~\bibnamefont {Vogt}}, \bibinfo {author}
  {\bibfnamefont {A.}~\bibnamefont {Al-Masoudi}}, \bibinfo {author}
  {\bibfnamefont {S.}~\bibnamefont {D\"orscher}}, \bibinfo {author}
  {\bibfnamefont {S.}~\bibnamefont {H\"afner}}, \bibinfo {author}
  {\bibfnamefont {U.}~\bibnamefont {Sterr}},\ and\ \bibinfo {author}
  {\bibfnamefont {C.}~\bibnamefont {Lisdat}},\ }\bibfield  {title} {\bibinfo
  {title} {Transportable optical lattice clock with $7\times{10}^{-17}$
  uncertainty},\ }\href {https://doi.org/10.1103/PhysRevLett.118.073601}
  {\bibfield  {journal} {\bibinfo  {journal} {Phys. Rev. Lett.}\ }\textbf
  {\bibinfo {volume} {118}},\ \bibinfo {pages} {073601} (\bibinfo {year}
  {2017})}\BibitemShut {NoStop}%
\bibitem [{\citenamefont {Huang}\ \emph {et~al.}(2020)\citenamefont {Huang},
  \citenamefont {Zhang}, \citenamefont {Zhang}, \citenamefont {Hao},
  \citenamefont {Guan}, \citenamefont {Zeng}, \citenamefont {Chen},
  \citenamefont {Lin}, \citenamefont {Wang}, \citenamefont {Cao}, \citenamefont
  {Liang}, \citenamefont {Fang}, \citenamefont {Fang}, \citenamefont {Li},\
  and\ \citenamefont {Gao}}]{Huang2020}%
  \BibitemOpen
  \bibfield  {author} {\bibinfo {author} {\bibfnamefont {Y.}~\bibnamefont
  {Huang}}, \bibinfo {author} {\bibfnamefont {H.}~\bibnamefont {Zhang}},
  \bibinfo {author} {\bibfnamefont {B.}~\bibnamefont {Zhang}}, \bibinfo
  {author} {\bibfnamefont {Y.}~\bibnamefont {Hao}}, \bibinfo {author}
  {\bibfnamefont {H.}~\bibnamefont {Guan}}, \bibinfo {author} {\bibfnamefont
  {M.}~\bibnamefont {Zeng}}, \bibinfo {author} {\bibfnamefont {Q.}~\bibnamefont
  {Chen}}, \bibinfo {author} {\bibfnamefont {Y.}~\bibnamefont {Lin}}, \bibinfo
  {author} {\bibfnamefont {Y.}~\bibnamefont {Wang}}, \bibinfo {author}
  {\bibfnamefont {S.}~\bibnamefont {Cao}}, \bibinfo {author} {\bibfnamefont
  {K.}~\bibnamefont {Liang}}, \bibinfo {author} {\bibfnamefont
  {F.}~\bibnamefont {Fang}}, \bibinfo {author} {\bibfnamefont {Z.}~\bibnamefont
  {Fang}}, \bibinfo {author} {\bibfnamefont {T.}~\bibnamefont {Li}},\ and\
  \bibinfo {author} {\bibfnamefont {K.}~\bibnamefont {Gao}},\ }\bibfield
  {title} {\bibinfo {title} {{Geopotential measurement with a robust,
  transportable Ca+ optical clock}},\ }\href
  {https://doi.org/10.1103/PhysRevA.102.050802} {\bibfield  {journal} {\bibinfo
   {journal} {Physical Review A}\ }\textbf {\bibinfo {volume} {102}},\ \bibinfo
  {pages} {050802} (\bibinfo {year} {2020})}\BibitemShut {NoStop}%
\bibitem [{\citenamefont {Wang}\ \emph {et~al.}(2020)\citenamefont {Wang},
  \citenamefont {Cao}, \citenamefont {Yuan}, \citenamefont {Liu}, \citenamefont
  {Shu},\ and\ \citenamefont {Huang}}]{Wang2020}%
  \BibitemOpen
  \bibfield  {author} {\bibinfo {author} {\bibfnamefont {S.}~\bibnamefont
  {Wang}}, \bibinfo {author} {\bibfnamefont {J.}~\bibnamefont {Cao}}, \bibinfo
  {author} {\bibfnamefont {J.}~\bibnamefont {Yuan}}, \bibinfo {author}
  {\bibfnamefont {D.}~\bibnamefont {Liu}}, \bibinfo {author} {\bibfnamefont
  {H.}~\bibnamefont {Shu}},\ and\ \bibinfo {author} {\bibfnamefont
  {X.}~\bibnamefont {Huang}},\ }\bibfield  {title} {\bibinfo {title}
  {{Integrated multiple wavelength stabilization on a multi-channel cavity for
  a transportable optical clock}},\ }\href {https://doi.org/10.1364/OE.383115}
  {\bibfield  {journal} {\bibinfo  {journal} {Optics Express}\ }\textbf
  {\bibinfo {volume} {28}},\ \bibinfo {pages} {11852} (\bibinfo {year}
  {2020})}\BibitemShut {NoStop}%
\bibitem [{\citenamefont {Abbasov}\ \emph {et~al.}(2020)\citenamefont
  {Abbasov}, \citenamefont {Makarenko}, \citenamefont {Sherstov}, \citenamefont
  {Axenov},\ and\ \citenamefont {Zalivako}}]{Abbasov2020}%
  \BibitemOpen
  \bibfield  {author} {\bibinfo {author} {\bibfnamefont {T.}~\bibnamefont
  {Abbasov}}, \bibinfo {author} {\bibfnamefont {K.}~\bibnamefont {Makarenko}},
  \bibinfo {author} {\bibfnamefont {I.}~\bibnamefont {Sherstov}}, \bibinfo
  {author} {\bibfnamefont {M.}~\bibnamefont {Axenov}},\ and\ \bibinfo {author}
  {\bibfnamefont {I.}~\bibnamefont {Zalivako}},\ }\bibfield  {title} {\bibinfo
  {title} {{Compact transportable $^{171}Yb^+$ single-ion optical fully
  automated clock with $4.9\times10^{-16}$ relative instability}},\ }\href@noop
  {} {\bibfield  {journal} {\bibinfo  {journal} {arXiv}\ ,\ \bibinfo {pages}
  {2}} (\bibinfo {year} {2020})}\BibitemShut {NoStop}%
\bibitem [{\citenamefont {Cao}\ \emph {et~al.}(2022)\citenamefont {Cao},
  \citenamefont {Yuan}, \citenamefont {Wang}, \citenamefont {Zhang},
  \citenamefont {Yuan}, \citenamefont {Liu}, \citenamefont {Cui}, \citenamefont
  {Chao}, \citenamefont {Shu}, \citenamefont {Lin}, \citenamefont {Cao},
  \citenamefont {Wang}, \citenamefont {Fang}, \citenamefont {Fang},
  \citenamefont {Li},\ and\ \citenamefont {Huang}}]{Cao2022}%
  \BibitemOpen
  \bibfield  {author} {\bibinfo {author} {\bibfnamefont {J.}~\bibnamefont
  {Cao}}, \bibinfo {author} {\bibfnamefont {J.}~\bibnamefont {Yuan}}, \bibinfo
  {author} {\bibfnamefont {S.}~\bibnamefont {Wang}}, \bibinfo {author}
  {\bibfnamefont {P.}~\bibnamefont {Zhang}}, \bibinfo {author} {\bibfnamefont
  {Y.}~\bibnamefont {Yuan}}, \bibinfo {author} {\bibfnamefont {D.}~\bibnamefont
  {Liu}}, \bibinfo {author} {\bibfnamefont {K.}~\bibnamefont {Cui}}, \bibinfo
  {author} {\bibfnamefont {S.}~\bibnamefont {Chao}}, \bibinfo {author}
  {\bibfnamefont {H.}~\bibnamefont {Shu}}, \bibinfo {author} {\bibfnamefont
  {Y.}~\bibnamefont {Lin}}, \bibinfo {author} {\bibfnamefont {S.}~\bibnamefont
  {Cao}}, \bibinfo {author} {\bibfnamefont {Y.}~\bibnamefont {Wang}}, \bibinfo
  {author} {\bibfnamefont {Z.}~\bibnamefont {Fang}}, \bibinfo {author}
  {\bibfnamefont {F.}~\bibnamefont {Fang}}, \bibinfo {author} {\bibfnamefont
  {T.}~\bibnamefont {Li}},\ and\ \bibinfo {author} {\bibfnamefont
  {X.}~\bibnamefont {Huang}},\ }\bibfield  {title} {\bibinfo {title} {{A
  compact, transportable optical clock with $1\times10^{-17}$ uncertainty and
  its absolute frequency measurement}},\ }\href
  {https://doi.org/10.1063/5.0079432} {\bibfield  {journal} {\bibinfo
  {journal} {Applied Physics Letters}\ }\textbf {\bibinfo {volume} {120}},\
  \bibinfo {pages} {1} (\bibinfo {year} {2022})}\BibitemShut {NoStop}%
\bibitem [{\citenamefont {Döringshoff}\ \emph {et~al.}(2017)\citenamefont
  {Döringshoff}, \citenamefont {Schuldt}, \citenamefont {Kovalchuk},
  \citenamefont {Stühler}, \citenamefont {Braxmaier},\ and\ \citenamefont
  {Peters}}]{Doringshoff2017}%
  \BibitemOpen
  \bibfield  {author} {\bibinfo {author} {\bibfnamefont {K.}~\bibnamefont
  {Döringshoff}}, \bibinfo {author} {\bibfnamefont {T.}~\bibnamefont
  {Schuldt}}, \bibinfo {author} {\bibfnamefont {E.}~\bibnamefont {Kovalchuk}},
  \bibinfo {author} {\bibfnamefont {J.}~\bibnamefont {Stühler}}, \bibinfo
  {author} {\bibfnamefont {C.}~\bibnamefont {Braxmaier}},\ and\ \bibinfo
  {author} {\bibfnamefont {A.}~\bibnamefont {Peters}},\ }\bibfield  {title}
  {\bibinfo {title} {A flight-like absolute optical frequency reference based
  on iodine for laser systems at 1064 nm},\ }\href
  {https://doi.org/10.1007/s00340-017-6756-1} {\bibfield  {journal} {\bibinfo
  {journal} {Applied Physics B: Lasers and Optics}\ }\textbf {\bibinfo {volume}
  {123}} (\bibinfo {year} {2017})}\BibitemShut {NoStop}%
\bibitem [{\citenamefont {D\"oringshoff}\ \emph {et~al.}(2019)\citenamefont
  {D\"oringshoff}, \citenamefont {Gutsch}, \citenamefont {Schkolnik},
  \citenamefont {K\"urbis}, \citenamefont {Oswald}, \citenamefont {Pr\"obster},
  \citenamefont {Kovalchuk}, \citenamefont {Bawamia}, \citenamefont {Smol},
  \citenamefont {Schuldt}, \citenamefont {Lezius}, \citenamefont {Holzwarth},
  \citenamefont {Wicht}, \citenamefont {Braxmaier}, \citenamefont {Krutzik},\
  and\ \citenamefont {Peters}}]{Doringshoff2019}%
  \BibitemOpen
  \bibfield  {author} {\bibinfo {author} {\bibfnamefont {K.}~\bibnamefont
  {D\"oringshoff}}, \bibinfo {author} {\bibfnamefont {F.~B.}\ \bibnamefont
  {Gutsch}}, \bibinfo {author} {\bibfnamefont {V.}~\bibnamefont {Schkolnik}},
  \bibinfo {author} {\bibfnamefont {C.}~\bibnamefont {K\"urbis}}, \bibinfo
  {author} {\bibfnamefont {M.}~\bibnamefont {Oswald}}, \bibinfo {author}
  {\bibfnamefont {B.}~\bibnamefont {Pr\"obster}}, \bibinfo {author}
  {\bibfnamefont {E.~V.}\ \bibnamefont {Kovalchuk}}, \bibinfo {author}
  {\bibfnamefont {A.}~\bibnamefont {Bawamia}}, \bibinfo {author} {\bibfnamefont
  {R.}~\bibnamefont {Smol}}, \bibinfo {author} {\bibfnamefont {T.}~\bibnamefont
  {Schuldt}}, \bibinfo {author} {\bibfnamefont {M.}~\bibnamefont {Lezius}},
  \bibinfo {author} {\bibfnamefont {R.}~\bibnamefont {Holzwarth}}, \bibinfo
  {author} {\bibfnamefont {A.}~\bibnamefont {Wicht}}, \bibinfo {author}
  {\bibfnamefont {C.}~\bibnamefont {Braxmaier}}, \bibinfo {author}
  {\bibfnamefont {M.}~\bibnamefont {Krutzik}},\ and\ \bibinfo {author}
  {\bibfnamefont {A.}~\bibnamefont {Peters}},\ }\bibfield  {title} {\bibinfo
  {title} {Iodine frequency reference on a sounding rocket},\ }\href
  {https://doi.org/10.1103/PhysRevApplied.11.054068} {\bibfield  {journal}
  {\bibinfo  {journal} {Phys. Rev. Appl.}\ }\textbf {\bibinfo {volume} {11}},\
  \bibinfo {pages} {054068} (\bibinfo {year} {2019})}\BibitemShut {NoStop}%
\bibitem [{\citenamefont {Schuldt}\ \emph {et~al.}(2017)\citenamefont
  {Schuldt}, \citenamefont {D\"{o}ringshoff}, \citenamefont {Kovalchuk},
  \citenamefont {Keetman}, \citenamefont {Pahl}, \citenamefont {Peters},\ and\
  \citenamefont {Braxmaier}}]{Schuldt2017}%
  \BibitemOpen
  \bibfield  {author} {\bibinfo {author} {\bibfnamefont {T.}~\bibnamefont
  {Schuldt}}, \bibinfo {author} {\bibfnamefont {K.}~\bibnamefont
  {D\"{o}ringshoff}}, \bibinfo {author} {\bibfnamefont {E.~V.}\ \bibnamefont
  {Kovalchuk}}, \bibinfo {author} {\bibfnamefont {A.}~\bibnamefont {Keetman}},
  \bibinfo {author} {\bibfnamefont {J.}~\bibnamefont {Pahl}}, \bibinfo {author}
  {\bibfnamefont {A.}~\bibnamefont {Peters}},\ and\ \bibinfo {author}
  {\bibfnamefont {C.}~\bibnamefont {Braxmaier}},\ }\bibfield  {title} {\bibinfo
  {title} {Development of a compact optical absolute frequency reference for
  space with 10{\textminus}15 instability},\ }\href
  {https://doi.org/10.1364/AO.56.001101} {\bibfield  {journal} {\bibinfo
  {journal} {Appl. Opt.}\ }\textbf {\bibinfo {volume} {56}},\ \bibinfo {pages}
  {1101} (\bibinfo {year} {2017})}\BibitemShut {NoStop}%
\bibitem [{\citenamefont {Nez}\ \emph {et~al.}(1993)\citenamefont {Nez},
  \citenamefont {Biraben}, \citenamefont {Felder},\ and\ \citenamefont
  {Millerioux}}]{Nez1993}%
  \BibitemOpen
  \bibfield  {author} {\bibinfo {author} {\bibfnamefont {F.}~\bibnamefont
  {Nez}}, \bibinfo {author} {\bibfnamefont {F.}~\bibnamefont {Biraben}},
  \bibinfo {author} {\bibfnamefont {R.}~\bibnamefont {Felder}},\ and\ \bibinfo
  {author} {\bibfnamefont {Y.}~\bibnamefont {Millerioux}},\ }\bibfield  {title}
  {\bibinfo {title} {Optical frequency determination of the hyperfine
  components of the {5S12}-{5D32} two-photon transitions in rubidium},\ }\href
  {https://doi.org/10.1016/0030-4018(93)90417-4} {\bibfield  {journal}
  {\bibinfo  {journal} {Optics Communications}\ }\textbf {\bibinfo {volume}
  {102}},\ \bibinfo {pages} {432} (\bibinfo {year} {1993})}\BibitemShut
  {NoStop}%
\bibitem [{\citenamefont {Maurice}\ \emph {et~al.}(2020)\citenamefont
  {Maurice}, \citenamefont {Newman}, \citenamefont {Dickerson}, \citenamefont
  {Rivers}, \citenamefont {Hsiao}, \citenamefont {Greene}, \citenamefont
  {Mescher}, \citenamefont {Kitching}, \citenamefont {Hummon},\ and\
  \citenamefont {Johnson}}]{Maurice2020}%
  \BibitemOpen
  \bibfield  {author} {\bibinfo {author} {\bibfnamefont {V.}~\bibnamefont
  {Maurice}}, \bibinfo {author} {\bibfnamefont {Z.~L.}\ \bibnamefont {Newman}},
  \bibinfo {author} {\bibfnamefont {S.}~\bibnamefont {Dickerson}}, \bibinfo
  {author} {\bibfnamefont {M.}~\bibnamefont {Rivers}}, \bibinfo {author}
  {\bibfnamefont {J.}~\bibnamefont {Hsiao}}, \bibinfo {author} {\bibfnamefont
  {P.}~\bibnamefont {Greene}}, \bibinfo {author} {\bibfnamefont
  {M.}~\bibnamefont {Mescher}}, \bibinfo {author} {\bibfnamefont
  {J.}~\bibnamefont {Kitching}}, \bibinfo {author} {\bibfnamefont {M.~T.}\
  \bibnamefont {Hummon}},\ and\ \bibinfo {author} {\bibfnamefont
  {C.}~\bibnamefont {Johnson}},\ }\bibfield  {title} {\bibinfo {title}
  {{Miniaturized optical frequency reference for next-generation portable
  optical clocks}},\ }\href {https://doi.org/10.1364/oe.396296} {\bibfield
  {journal} {\bibinfo  {journal} {Optics Express}\ }\textbf {\bibinfo {volume}
  {28}},\ \bibinfo {pages} {24708} (\bibinfo {year} {2020})}\BibitemShut
  {NoStop}%
\bibitem [{\citenamefont {Newman}\ \emph {et~al.}(2021)\citenamefont {Newman},
  \citenamefont {Maurice}, \citenamefont {Fredrick}, \citenamefont {Fortier},
  \citenamefont {Leopardi}, \citenamefont {Hollberg}, \citenamefont {Diddams},
  \citenamefont {Kitching},\ and\ \citenamefont {Hummon}}]{Newman2021}%
  \BibitemOpen
  \bibfield  {author} {\bibinfo {author} {\bibfnamefont {Z.~L.}\ \bibnamefont
  {Newman}}, \bibinfo {author} {\bibfnamefont {V.}~\bibnamefont {Maurice}},
  \bibinfo {author} {\bibfnamefont {C.}~\bibnamefont {Fredrick}}, \bibinfo
  {author} {\bibfnamefont {T.}~\bibnamefont {Fortier}}, \bibinfo {author}
  {\bibfnamefont {H.}~\bibnamefont {Leopardi}}, \bibinfo {author}
  {\bibfnamefont {L.}~\bibnamefont {Hollberg}}, \bibinfo {author}
  {\bibfnamefont {S.~A.}\ \bibnamefont {Diddams}}, \bibinfo {author}
  {\bibfnamefont {J.}~\bibnamefont {Kitching}},\ and\ \bibinfo {author}
  {\bibfnamefont {M.~T.}\ \bibnamefont {Hummon}},\ }\bibfield  {title}
  {\bibinfo {title} {High-performance, compact optical standard},\ }\href
  {https://doi.org/10.1364/OL.435603} {\bibfield  {journal} {\bibinfo
  {journal} {Opt. Lett.}\ }\textbf {\bibinfo {volume} {46}},\ \bibinfo {pages}
  {4702} (\bibinfo {year} {2021})}\BibitemShut {NoStop}%
\bibitem [{\citenamefont {Lemke}\ \emph {et~al.}(2022)\citenamefont {Lemke},
  \citenamefont {Martin}, \citenamefont {Beard}, \citenamefont {Stuhl},
  \citenamefont {Metcalf},\ and\ \citenamefont {Elgin}}]{Lemke2022}%
  \BibitemOpen
  \bibfield  {author} {\bibinfo {author} {\bibfnamefont {N.~D.}\ \bibnamefont
  {Lemke}}, \bibinfo {author} {\bibfnamefont {K.~W.}\ \bibnamefont {Martin}},
  \bibinfo {author} {\bibfnamefont {R.}~\bibnamefont {Beard}}, \bibinfo
  {author} {\bibfnamefont {B.~K.}\ \bibnamefont {Stuhl}}, \bibinfo {author}
  {\bibfnamefont {A.~J.}\ \bibnamefont {Metcalf}},\ and\ \bibinfo {author}
  {\bibfnamefont {J.~D.}\ \bibnamefont {Elgin}},\ }\bibfield  {title} {\bibinfo
  {title} {Measurement of optical rubidium clock frequency spanning 65 days},\
  }\bibfield  {journal} {\bibinfo  {journal} {Sensors}\ }\textbf {\bibinfo
  {volume} {22}},\ \href {https://doi.org/10.3390/s22051982}
  {10.3390/s22051982} (\bibinfo {year} {2022})\BibitemShut {NoStop}%
\bibitem [{\citenamefont {Perrella}\ \emph {et~al.}(2019)\citenamefont
  {Perrella}, \citenamefont {Light}, \citenamefont {Anstie}, \citenamefont
  {Baynes}, \citenamefont {White},\ and\ \citenamefont
  {Luiten}}]{Perrella2019}%
  \BibitemOpen
  \bibfield  {author} {\bibinfo {author} {\bibfnamefont {C.}~\bibnamefont
  {Perrella}}, \bibinfo {author} {\bibfnamefont {P.~S.}\ \bibnamefont {Light}},
  \bibinfo {author} {\bibfnamefont {J.~D.}\ \bibnamefont {Anstie}}, \bibinfo
  {author} {\bibfnamefont {F.~N.}\ \bibnamefont {Baynes}}, \bibinfo {author}
  {\bibfnamefont {R.~T.}\ \bibnamefont {White}},\ and\ \bibinfo {author}
  {\bibfnamefont {A.~N.}\ \bibnamefont {Luiten}},\ }\bibfield  {title}
  {\bibinfo {title} {{Dichroic Two-Photon Rubidium Frequency Standard}},\
  }\href {https://doi.org/10.1103/PhysRevApplied.12.054063} {\bibfield
  {journal} {\bibinfo  {journal} {Physical Review Applied}\ }\textbf {\bibinfo
  {volume} {12}},\ \bibinfo {pages} {1} (\bibinfo {year} {2019})}\BibitemShut
  {NoStop}%
\bibitem [{\citenamefont {Bjorkholm}\ and\ \citenamefont
  {Liao}(1976)}]{Bjorkholm1976}%
  \BibitemOpen
  \bibfield  {author} {\bibinfo {author} {\bibfnamefont {J.~E.}\ \bibnamefont
  {Bjorkholm}}\ and\ \bibinfo {author} {\bibfnamefont {P.~F.}\ \bibnamefont
  {Liao}},\ }\bibfield  {title} {\bibinfo {title} {Line shape and strength of
  two-photon absorption in an atomic vapor with a resonant or nearly resonant
  intermediate state},\ }\href {https://doi.org/10.1103/PhysRevA.14.751}
  {\bibfield  {journal} {\bibinfo  {journal} {Physical Review A}\ }\textbf
  {\bibinfo {volume} {14}},\ \bibinfo {pages} {751} (\bibinfo {year} {1976})},\
  \bibinfo {note} {publisher: American Physical Society}\BibitemShut {NoStop}%
\bibitem [{\citenamefont {Gerginov}\ and\ \citenamefont
  {Beloy}(2018)}]{Gerginov2018}%
  \BibitemOpen
  \bibfield  {author} {\bibinfo {author} {\bibfnamefont {V.}~\bibnamefont
  {Gerginov}}\ and\ \bibinfo {author} {\bibfnamefont {K.}~\bibnamefont
  {Beloy}},\ }\bibfield  {title} {\bibinfo {title} {Two-photon optical
  frequency reference with active ac stark shift cancellation},\ }\href
  {https://doi.org/10.1103/PhysRevApplied.10.014031} {\bibfield  {journal}
  {\bibinfo  {journal} {Phys. Rev. Appl.}\ }\textbf {\bibinfo {volume} {10}},\
  \bibinfo {pages} {014031} (\bibinfo {year} {2018})}\BibitemShut {NoStop}%
\bibitem [{\citenamefont {Hamilton}\ \emph {et~al.}(2023)\citenamefont
  {Hamilton}, \citenamefont {Roberts}, \citenamefont {Scholten}, \citenamefont
  {Locke}, \citenamefont {Luiten}, \citenamefont {Ginges},\ and\ \citenamefont
  {Perrella}}]{Hamilton2023}%
  \BibitemOpen
  \bibfield  {author} {\bibinfo {author} {\bibfnamefont {R.}~\bibnamefont
  {Hamilton}}, \bibinfo {author} {\bibfnamefont {B.~M.}\ \bibnamefont
  {Roberts}}, \bibinfo {author} {\bibfnamefont {S.~K.}\ \bibnamefont
  {Scholten}}, \bibinfo {author} {\bibfnamefont {C.}~\bibnamefont {Locke}},
  \bibinfo {author} {\bibfnamefont {A.~N.}\ \bibnamefont {Luiten}}, \bibinfo
  {author} {\bibfnamefont {J.~S.}\ \bibnamefont {Ginges}},\ and\ \bibinfo
  {author} {\bibfnamefont {C.}~\bibnamefont {Perrella}},\ }\bibfield  {title}
  {\bibinfo {title} {Experimental and theoretical study of dynamic
  polarizabilities in the $5{S}_{1/2}$--$5{D}_{5/2}$ clock transition in
  rubidium-87 and determination of electric dipole matrix elements},\ }\href
  {https://doi.org/10.1103/PhysRevApplied.19.054059} {\bibfield  {journal}
  {\bibinfo  {journal} {Phys. Rev. Appl.}\ }\textbf {\bibinfo {volume} {19}},\
  \bibinfo {pages} {054059} (\bibinfo {year} {2023})}\BibitemShut {NoStop}%
\bibitem [{\citenamefont {Nguyen}\ and\ \citenamefont
  {Schibli}(2022)}]{Nguyen2022}%
  \BibitemOpen
  \bibfield  {author} {\bibinfo {author} {\bibfnamefont {T.~N.}\ \bibnamefont
  {Nguyen}}\ and\ \bibinfo {author} {\bibfnamefont {T.~R.}\ \bibnamefont
  {Schibli}},\ }\bibfield  {title} {\bibinfo {title}
  {Temperature-shift-suppression scheme for two-photon two-color rubidium vapor
  clocks},\ }\href {https://doi.org/10.1103/PhysRevA.106.053104} {\bibfield
  {journal} {\bibinfo  {journal} {Phys. Rev. A}\ }\textbf {\bibinfo {volume}
  {106}},\ \bibinfo {pages} {053104} (\bibinfo {year} {2022})}\BibitemShut
  {NoStop}%
\bibitem [{\citenamefont {Margolis}(2010)}]{Margolis2010}%
  \BibitemOpen
  \bibfield  {author} {\bibinfo {author} {\bibfnamefont {H.~S.}\ \bibnamefont
  {Margolis}},\ }\bibfield  {title} {\bibinfo {title} {Optical frequency
  standards and clocks},\ }\href {https://doi.org/10.1080/00107510903257616}
  {\bibfield  {journal} {\bibinfo  {journal} {Contemporary Physics}\ }\textbf
  {\bibinfo {volume} {51}},\ \bibinfo {pages} {37} (\bibinfo {year}
  {2010})}\BibitemShut {NoStop}%
\bibitem [{\citenamefont {Sheng}\ \emph {et~al.}(2008)\citenamefont {Sheng},
  \citenamefont {Galvan},\ and\ \citenamefont {Orozco}}]{Sheng2008}%
  \BibitemOpen
  \bibfield  {author} {\bibinfo {author} {\bibfnamefont {D.}~\bibnamefont
  {Sheng}}, \bibinfo {author} {\bibfnamefont {A.}~\bibnamefont {Galvan}},\ and\
  \bibinfo {author} {\bibfnamefont {L.}~\bibnamefont {Orozco}},\ }\bibfield
  {title} {\bibinfo {title} {Lifetime measurements of the 5d states of
  rubidium},\ }\href {https://doi.org/10.1103/PhysRevA.78.062506} {\bibfield
  {journal} {\bibinfo  {journal} {Physical Review A}\ }\textbf {\bibinfo
  {volume} {78}} (\bibinfo {year} {2008})}\BibitemShut {NoStop}%
\bibitem [{\citenamefont {Martin}\ \emph {et~al.}(2018)\citenamefont {Martin},
  \citenamefont {Phelps}, \citenamefont {Lemke}, \citenamefont {Bigelow},
  \citenamefont {Stuhl}, \citenamefont {Wojcik}, \citenamefont {Holt},
  \citenamefont {Coddington}, \citenamefont {Bishop},\ and\ \citenamefont
  {Burke}}]{Martin2018}%
  \BibitemOpen
  \bibfield  {author} {\bibinfo {author} {\bibfnamefont {K.~W.}\ \bibnamefont
  {Martin}}, \bibinfo {author} {\bibfnamefont {G.}~\bibnamefont {Phelps}},
  \bibinfo {author} {\bibfnamefont {N.~D.}\ \bibnamefont {Lemke}}, \bibinfo
  {author} {\bibfnamefont {M.~S.}\ \bibnamefont {Bigelow}}, \bibinfo {author}
  {\bibfnamefont {B.}~\bibnamefont {Stuhl}}, \bibinfo {author} {\bibfnamefont
  {M.}~\bibnamefont {Wojcik}}, \bibinfo {author} {\bibfnamefont
  {M.}~\bibnamefont {Holt}}, \bibinfo {author} {\bibfnamefont {I.}~\bibnamefont
  {Coddington}}, \bibinfo {author} {\bibfnamefont {M.~W.}\ \bibnamefont
  {Bishop}},\ and\ \bibinfo {author} {\bibfnamefont {J.~H.}\ \bibnamefont
  {Burke}},\ }\bibfield  {title} {\bibinfo {title} {Compact optical atomic
  clock based on a two-photon transition in rubidium},\ }\href
  {https://doi.org/10.1103/PhysRevApplied.9.014019} {\bibfield  {journal}
  {\bibinfo  {journal} {Phys. Rev. Appl.}\ }\textbf {\bibinfo {volume} {9}},\
  \bibinfo {pages} {014019} (\bibinfo {year} {2018})}\BibitemShut {NoStop}%
\bibitem [{\citenamefont {Safronova}\ \emph {et~al.}(2004)\citenamefont
  {Safronova}, \citenamefont {Williams},\ and\ \citenamefont
  {Clark}}]{Safronova2004}%
  \BibitemOpen
  \bibfield  {author} {\bibinfo {author} {\bibfnamefont {M.~S.}\ \bibnamefont
  {Safronova}}, \bibinfo {author} {\bibfnamefont {C.~J.}\ \bibnamefont
  {Williams}},\ and\ \bibinfo {author} {\bibfnamefont {C.~W.}\ \bibnamefont
  {Clark}},\ }\bibfield  {title} {\bibinfo {title} {Relativistic many-body
  calculations of electric-dipole matrix elements, lifetimes, and
  polarizabilities in rubidium},\ }\href
  {https://doi.org/10.1103/PhysRevA.69.022509} {\bibfield  {journal} {\bibinfo
  {journal} {Physical Review A - Atomic, Molecular, and Optical Physics}\
  }\textbf {\bibinfo {volume} {69}},\ \bibinfo {pages} {8} (\bibinfo {year}
  {2004})}\BibitemShut {NoStop}%
\bibitem [{\citenamefont {Hassanin}\ \emph {et~al.}(2023)\citenamefont
  {Hassanin}, \citenamefont {Federsel}, \citenamefont {Karlewski},\ and\
  \citenamefont {Zimmermann}}]{Hassanin2023}%
  \BibitemOpen
  \bibfield  {author} {\bibinfo {author} {\bibfnamefont {K.}~\bibnamefont
  {Hassanin}}, \bibinfo {author} {\bibfnamefont {P.}~\bibnamefont {Federsel}},
  \bibinfo {author} {\bibfnamefont {F.}~\bibnamefont {Karlewski}},\ and\
  \bibinfo {author} {\bibfnamefont {C.}~\bibnamefont {Zimmermann}},\ }\bibfield
   {title} {\bibinfo {title}
  {\${5S}{\textbackslash}text\{{\textbackslash}ensuremath\{-\}\}{5D}\$
  two-photon transition in rubidium vapor at high densities},\ }\href
  {https://doi.org/10.1103/PhysRevA.107.043104} {\bibfield  {journal} {\bibinfo
   {journal} {Physical Review A}\ }\textbf {\bibinfo {volume} {107}},\ \bibinfo
  {pages} {043104} (\bibinfo {year} {2023})}\BibitemShut {NoStop}%
\bibitem [{\citenamefont {Sinclair}\ \emph {et~al.}(2015)\citenamefont
  {Sinclair}, \citenamefont {Desch{\^e}nes}, \citenamefont {Sonderhouse},
  \citenamefont {Swann}, \citenamefont {Khader}, \citenamefont {Baumann},
  \citenamefont {Newbury},\ and\ \citenamefont
  {Coddington}}]{Sinclair2015InvitedAA}%
  \BibitemOpen
  \bibfield  {author} {\bibinfo {author} {\bibfnamefont {L.~C.}\ \bibnamefont
  {Sinclair}}, \bibinfo {author} {\bibfnamefont {J.-D.}\ \bibnamefont
  {Desch{\^e}nes}}, \bibinfo {author} {\bibfnamefont {L.}~\bibnamefont
  {Sonderhouse}}, \bibinfo {author} {\bibfnamefont {W.~C.}\ \bibnamefont
  {Swann}}, \bibinfo {author} {\bibfnamefont {I.}~\bibnamefont {Khader}},
  \bibinfo {author} {\bibfnamefont {E.}~\bibnamefont {Baumann}}, \bibinfo
  {author} {\bibfnamefont {N.~R.}\ \bibnamefont {Newbury}},\ and\ \bibinfo
  {author} {\bibfnamefont {I.}~\bibnamefont {Coddington}},\ }\bibfield  {title}
  {\bibinfo {title} {Invited article: A compact optically coherent fiber
  frequency comb.},\ }\href@noop {} {\bibfield  {journal} {\bibinfo  {journal}
  {The Review of scientific instruments}\ }\textbf {\bibinfo {volume} {86 8}},\
  \bibinfo {pages} {081301} (\bibinfo {year} {2015})}\BibitemShut {NoStop}%
\bibitem [{\citenamefont {Tobar}\ \emph {et~al.}(2006)\citenamefont {Tobar},
  \citenamefont {Ivanov}, \citenamefont {Locke}, \citenamefont {Stanwix},
  \citenamefont {Hartnett}, \citenamefont {Luiten}, \citenamefont {Warrington},
  \citenamefont {H.~Fisk}, \citenamefont {Lawn}, \citenamefont {Wouters},
  \citenamefont {Bize}, \citenamefont {Santarelli}, \citenamefont {Wolf},
  \citenamefont {Clairon},\ and\ \citenamefont {Guillemot}}]{Tobar2006}%
  \BibitemOpen
  \bibfield  {author} {\bibinfo {author} {\bibfnamefont {M.~E.}\ \bibnamefont
  {Tobar}}, \bibinfo {author} {\bibfnamefont {E.~N.}\ \bibnamefont {Ivanov}},
  \bibinfo {author} {\bibfnamefont {C.~R.}\ \bibnamefont {Locke}}, \bibinfo
  {author} {\bibfnamefont {P.~L.}\ \bibnamefont {Stanwix}}, \bibinfo {author}
  {\bibfnamefont {J.~G.}\ \bibnamefont {Hartnett}}, \bibinfo {author}
  {\bibfnamefont {A.~N.}\ \bibnamefont {Luiten}}, \bibinfo {author}
  {\bibfnamefont {R.~B.}\ \bibnamefont {Warrington}}, \bibinfo {author}
  {\bibfnamefont {P.~T.}\ \bibnamefont {H.~Fisk}}, \bibinfo {author}
  {\bibfnamefont {M.~A.}\ \bibnamefont {Lawn}}, \bibinfo {author}
  {\bibfnamefont {M.~J.}\ \bibnamefont {Wouters}}, \bibinfo {author}
  {\bibfnamefont {S.}~\bibnamefont {Bize}}, \bibinfo {author} {\bibfnamefont
  {G.}~\bibnamefont {Santarelli}}, \bibinfo {author} {\bibfnamefont
  {P.}~\bibnamefont {Wolf}}, \bibinfo {author} {\bibfnamefont {A.}~\bibnamefont
  {Clairon}},\ and\ \bibinfo {author} {\bibfnamefont {P.}~\bibnamefont
  {Guillemot}},\ }\bibfield  {title} {\bibinfo {title} {Long-term operation and
  performance of cryogenic sapphire oscillators},\ }\href
  {https://doi.org/10.1109/TUFFC.2006.187} {\bibfield  {journal} {\bibinfo
  {journal} {IEEE Transactions on Ultrasonics, Ferroelectrics, and Frequency
  Control}\ }\textbf {\bibinfo {volume} {53}},\ \bibinfo {pages} {2386}
  (\bibinfo {year} {2006})}\BibitemShut {NoStop}%
\bibitem [{\citenamefont {Steck}(2023)}]{Steck2023}%
  \BibitemOpen
  \bibfield  {author} {\bibinfo {author} {\bibfnamefont {D.~A.}\ \bibnamefont
  {Steck}},\ }\href@noop {} {\bibinfo {title} {Rubidium 87 {D} {Line}
  {Data}}},\ \bibinfo {howpublished} {\url{http://steck.us/alkalidata}}
  (\bibinfo {year} {2023})\BibitemShut {NoStop}%
\bibitem [{\citenamefont {Rutman}\ and\ \citenamefont
  {Walls}(1991)}]{Rutman1991}%
  \BibitemOpen
  \bibfield  {author} {\bibinfo {author} {\bibfnamefont {J.}~\bibnamefont
  {Rutman}}\ and\ \bibinfo {author} {\bibfnamefont {F.}~\bibnamefont {Walls}},\
  }\bibfield  {title} {\bibinfo {title} {Characterization of {Frequency}
  {Stability} in {Precision} {Frequency} {Sources}},\ }\href
  {https://doi.org/10.1109/5.84972} {\bibfield  {journal} {\bibinfo  {journal}
  {Proceedings of the IEEE}\ }\textbf {\bibinfo {volume} {79}},\ \bibinfo
  {pages} {952} (\bibinfo {year} {1991})}\BibitemShut {NoStop}%
\bibitem [{\citenamefont {Turner}\ \emph {et~al.}(2002)\citenamefont {Turner},
  \citenamefont {Weber}, \citenamefont {Hawthorn},\ and\ \citenamefont
  {Scholten}}]{Turner2002}%
  \BibitemOpen
  \bibfield  {author} {\bibinfo {author} {\bibfnamefont {L.}~\bibnamefont
  {Turner}}, \bibinfo {author} {\bibfnamefont {K.}~\bibnamefont {Weber}},
  \bibinfo {author} {\bibfnamefont {C.}~\bibnamefont {Hawthorn}},\ and\
  \bibinfo {author} {\bibfnamefont {R.}~\bibnamefont {Scholten}},\ }\bibfield
  {title} {\bibinfo {title} {Frequency noise characterisation of narrow
  linewidth diode lasers},\ }\href
  {https://doi.org/https://doi.org/10.1016/S0030-4018(01)01689-3} {\bibfield
  {journal} {\bibinfo  {journal} {Optics Communications}\ }\textbf {\bibinfo
  {volume} {201}},\ \bibinfo {pages} {391} (\bibinfo {year}
  {2002})}\BibitemShut {NoStop}%
\bibitem [{\citenamefont {Audoin}\ \emph {et~al.}(1991)\citenamefont {Audoin},
  \citenamefont {Candelier},\ and\ \citenamefont {Diamarcq}}]{Audoin1991}%
  \BibitemOpen
  \bibfield  {author} {\bibinfo {author} {\bibfnamefont {C.}~\bibnamefont
  {Audoin}}, \bibinfo {author} {\bibfnamefont {V.}~\bibnamefont {Candelier}},\
  and\ \bibinfo {author} {\bibfnamefont {N.}~\bibnamefont {Diamarcq}},\
  }\bibfield  {title} {\bibinfo {title} {A limit to the frequency stability of
  passive frequency standards due to an intermodulation effect},\ }\href
  {https://doi.org/10.1109/TIM.1990.1032896} {\bibfield  {journal} {\bibinfo
  {journal} {IEEE Transactions on Instrumentation and Measurement}\ }\textbf
  {\bibinfo {volume} {40}},\ \bibinfo {pages} {121} (\bibinfo {year}
  {1991})}\BibitemShut {NoStop}%
\bibitem [{\citenamefont {Mitroy}\ \emph {et~al.}(2010)\citenamefont {Mitroy},
  \citenamefont {Safronova},\ and\ \citenamefont {Clark}}]{Mitroy2010}%
  \BibitemOpen
  \bibfield  {author} {\bibinfo {author} {\bibfnamefont {J.}~\bibnamefont
  {Mitroy}}, \bibinfo {author} {\bibfnamefont {M.}~\bibnamefont {Safronova}},\
  and\ \bibinfo {author} {\bibfnamefont {C.}~\bibnamefont {Clark}},\ }\bibfield
   {title} {\bibinfo {title} {Theory and applications of atomic and ionic
  polarizabilities},\ }\href {https://doi.org/10.1088/0953-4075/43/20/202001}
  {\bibfield  {journal} {\bibinfo  {journal} {Journal of Physics B: Atomic,
  Molecular and Optical Physics}\ }\textbf {\bibinfo {volume} {43}},\ \bibinfo
  {pages} {202001} (\bibinfo {year} {2010})}\BibitemShut {NoStop}%
\bibitem [{\citenamefont {Lewis}(1980)}]{LEWIS1980}%
  \BibitemOpen
  \bibfield  {author} {\bibinfo {author} {\bibfnamefont {E.}~\bibnamefont
  {Lewis}},\ }\bibfield  {title} {\bibinfo {title} {Collisional relaxation of
  atomic excited states, line broadening and interatomic interactions},\ }\href
  {https://doi.org/https://doi.org/10.1016/0370-1573(80)90056-3} {\bibfield
  {journal} {\bibinfo  {journal} {Physics Reports}\ }\textbf {\bibinfo {volume}
  {58}},\ \bibinfo {pages} {1} (\bibinfo {year} {1980})}\BibitemShut {NoStop}%
\bibitem [{\citenamefont {Edwards}(1973)}]{Edwards_1973}%
  \BibitemOpen
  \bibfield  {author} {\bibinfo {author} {\bibfnamefont {D.~M.}\ \bibnamefont
  {Edwards}},\ }\bibfield  {title} {\bibinfo {title} {Electronic transitions
  and the high pressure chemistry and physics of solids},\ }\href
  {https://doi.org/10.1088/0031-9112/24/11/034} {\bibfield  {journal} {\bibinfo
   {journal} {Physics Bulletin}\ }\textbf {\bibinfo {volume} {24}},\ \bibinfo
  {pages} {683} (\bibinfo {year} {1973})}\BibitemShut {NoStop}%
\bibitem [{\citenamefont {Perrella}\ \emph {et~al.}(2013)\citenamefont
  {Perrella}, \citenamefont {Light}, \citenamefont {Anstie}, \citenamefont
  {Baynes}, \citenamefont {Benabid},\ and\ \citenamefont
  {Luiten}}]{Perrella2013_HCF}%
  \BibitemOpen
  \bibfield  {author} {\bibinfo {author} {\bibfnamefont {C.}~\bibnamefont
  {Perrella}}, \bibinfo {author} {\bibfnamefont {P.~S.}\ \bibnamefont {Light}},
  \bibinfo {author} {\bibfnamefont {J.~D.}\ \bibnamefont {Anstie}}, \bibinfo
  {author} {\bibfnamefont {F.~N.}\ \bibnamefont {Baynes}}, \bibinfo {author}
  {\bibfnamefont {F.}~\bibnamefont {Benabid}},\ and\ \bibinfo {author}
  {\bibfnamefont {A.~N.}\ \bibnamefont {Luiten}},\ }\bibfield  {title}
  {\bibinfo {title} {Two-color rubidium fiber frequency standard},\ }\href
  {https://doi.org/10.1364/OL.38.002122} {\bibfield  {journal} {\bibinfo
  {journal} {Opt. Lett.}\ }\textbf {\bibinfo {volume} {38}},\ \bibinfo {pages}
  {2122} (\bibinfo {year} {2013})}\BibitemShut {NoStop}%
\bibitem [{\citenamefont {Sanner}\ \emph {et~al.}(2018)\citenamefont {Sanner},
  \citenamefont {Huntemann}, \citenamefont {Lange}, \citenamefont {Tamm},\ and\
  \citenamefont {Peik}}]{Sanner2018}%
  \BibitemOpen
  \bibfield  {author} {\bibinfo {author} {\bibfnamefont {C.}~\bibnamefont
  {Sanner}}, \bibinfo {author} {\bibfnamefont {N.}~\bibnamefont {Huntemann}},
  \bibinfo {author} {\bibfnamefont {R.}~\bibnamefont {Lange}}, \bibinfo
  {author} {\bibfnamefont {C.}~\bibnamefont {Tamm}},\ and\ \bibinfo {author}
  {\bibfnamefont {E.}~\bibnamefont {Peik}},\ }\bibfield  {title} {\bibinfo
  {title} {Autobalanced ramsey spectroscopy},\ }\href
  {https://doi.org/10.1103/PhysRevLett.120.053602} {\bibfield  {journal}
  {\bibinfo  {journal} {Phys. Rev. Lett.}\ }\textbf {\bibinfo {volume} {120}},\
  \bibinfo {pages} {053602} (\bibinfo {year} {2018})}\BibitemShut {NoStop}%
\bibitem [{\citenamefont {Yudin}\ \emph {et~al.}(2018)\citenamefont {Yudin},
  \citenamefont {Taichenachev}, \citenamefont {Basalaev}, \citenamefont
  {Zanon-Willette}, \citenamefont {Mehlstäubler}, \citenamefont {Boudot},
  \citenamefont {Pollock}, \citenamefont {Shuker}, \citenamefont {Donley},\
  and\ \citenamefont {Kitching}}]{Yudin2018}%
  \BibitemOpen
  \bibfield  {author} {\bibinfo {author} {\bibfnamefont {V.~I.}\ \bibnamefont
  {Yudin}}, \bibinfo {author} {\bibfnamefont {A.~V.}\ \bibnamefont
  {Taichenachev}}, \bibinfo {author} {\bibfnamefont {M.~Y.}\ \bibnamefont
  {Basalaev}}, \bibinfo {author} {\bibfnamefont {T.}~\bibnamefont
  {Zanon-Willette}}, \bibinfo {author} {\bibfnamefont {T.~E.}\ \bibnamefont
  {Mehlstäubler}}, \bibinfo {author} {\bibfnamefont {R.}~\bibnamefont
  {Boudot}}, \bibinfo {author} {\bibfnamefont {J.~W.}\ \bibnamefont {Pollock}},
  \bibinfo {author} {\bibfnamefont {M.}~\bibnamefont {Shuker}}, \bibinfo
  {author} {\bibfnamefont {E.~A.}\ \bibnamefont {Donley}},\ and\ \bibinfo
  {author} {\bibfnamefont {J.}~\bibnamefont {Kitching}},\ }\bibfield  {title}
  {\bibinfo {title} {Combined error signal in {Ramsey} spectroscopy of clock
  transitions},\ }\href {https://doi.org/10.1088/1367-2630/aaf47c} {\bibfield
  {journal} {\bibinfo  {journal} {New Journal of Physics}\ }\textbf {\bibinfo
  {volume} {20}},\ \bibinfo {pages} {123016} (\bibinfo {year}
  {2018})}\BibitemShut {NoStop}%
\end{thebibliography}


%

\end{document}